\documentclass{jfm_arxiv}
\usepackage{graphicx}
\usepackage{newtxtext}
\usepackage{newtxmath}
\usepackage{upgreek}  
\usepackage{natbib}
\usepackage[justification=justified,format=plain]{caption}
\usepackage{hyperref}
\hypersetup{
    colorlinks = true,
    urlcolor   = blue,
    citecolor  = blue,
}

\newcommand{\RomanNumeralCaps}[1]
\linenumbers

\usepackage{amsmath}
\usepackage{amsfonts}
\usepackage{bm}
\usepackage{bbm}
\usepackage{booktabs}

\newcommand{\bQ}{\ensuremath{\boldsymbol{Q}}}
\newcommand{\bq}{\ensuremath{\boldsymbol{q}}}
\newcommand{\bX}{\ensuremath{\boldsymbol{X}}}
\newcommand{\bbX}{\ensuremath{\boldsymbol{\bar{X}}}}
\newcommand{\bV}{\ensuremath{\boldsymbol{V}}}
\newcommand{\bbV}{\ensuremath{\boldsymbol{\bar{V}}}}
\newcommand{\bu}{\ensuremath{\boldsymbol{u}}}

\newcommand{\bn}{\ensuremath{\boldsymbol{n}}}
\newcommand{\bp}{\ensuremath{\boldsymbol{p}}}

\newcommand{\dd}{\ensuremath{\mathrm{d}}}

\newcommand{\bO}{\ensuremath{\boldsymbol{\Omega}}}
\newcommand{\bbu}{\ensuremath{\boldsymbol{\mathcal{U}}_\ell}}
\newcommand{\bba}{\ensuremath{\boldsymbol{\mathcal{G}}_\ell}}
\newcommand{\bg}{\ensuremath{\boldsymbol{g}}}
\newcommand{\St}{\mathrm{St}}

\title[Long rigid fibres in a turbulent channel flow]{Long rigid fibres in a turbulent channel flow: comparison between experiments and simulations}

\author[D. Sun \textit{et al.}]{Defa Sun\aff{1}, Florian Zumbo\aff{1}, Christophe Pitiot\aff{1}, Sandra Bosio\aff{1}, Cyrille Claudet\aff{1}, J\'er\'emie Bec\aff{1,2}, Christophe Brouzet\aff{1,2}\corresp{\email{christophe.brouzet@univ-cotedazur.fr}}}

\affiliation{\aff{1}Universit\'e C\^ote d'Azur, CNRS, INPHYNI, Nice, France
\aff{2}Universit\'{e} C\^{o}te d'Azur, Inria, CNRS, Calisto team, Sophia Antipolis, France}

\begin{document}
\maketitle

\begin{abstract}
The dynamics of long rigid fibres transported by turbulent channel flow are investigated experimentally and numerically. Experiments use polystyrene fibres of three lengths, $\ell/h=0.25$, $0.5$ and $1$, with moderate inertia, $St^+\approx20$. Their settling velocity is comparable to the friction velocity, causing accumulation near the bottom wall. Measurements are compared systematically with simulations based on a rigid slender-body model. Statistics conditioned on the distance from the wall are used to characterise the effects of fibre length and confinement on translation, orientation and tumbling. Away from the wall, the experimental fibres lag the fluid, with no clear dependence of the velocity deficit on length. Near the wall, the shortest fibres move faster than the local mean flow, whereas longer fibres remain slower, indicating length-dependent sampling of near-wall turbulence. Confinement also strongly constrains orientation and rotation. Fibres close to the wall predominantly undergo "kayaking" motion, tumbling in planes approximately parallel to it. Orientation statistics collapse when wall distance is normalised by fibre length, identifying $y^+/\ell^+$ as the relevant geometrical variable. Where fibres can acquire a significant wall-normal orientation, "pole-vaulting" events produce a local tumbling-rate maximum at $y^+\approx\ell^+/2$. Its magnitude decreases approximately as $(\ell^+)^{-2}$, consistently with a dimensional estimate based on the near-wall velocity variation sampled along the fibre. Experiments and simulations agree well for orientation and tumbling but differ more for translational velocity. The discrepancies highlight the effect of settling, finite fibre thickness and finite slip Reynolds number that are not fully represented by the model.
\end{abstract}



\section{Introduction}\label{sec:intro}

Turbulent flows laden with fibre-like particles are of paramount importance in various natural phenomena and industrial processes. On the environmental side, a few examples include aerosols in the atmosphere~\citep{NEWSOM1998}, the formation of ice crystals in cold clouds~\citep{gustav}, atmospheric and oceanic microplastic deposition or transport~\citep{allen:insu-02109784,dibenedetto}, and waste incineration~\citep{rn7}. On the industrial side, fibre suspensions are widely used in papermaking~\citep{annurevlundell}, can lead to drag reduction or drag increase~\citep{paschkewitz,marchioli20212021009}, depending on the orientation, volume fraction, aspect ratio, and flexibility, and can also clog water intake filters of chemical or energy plants~\citep{RedlingerPohnetal2022}. Given their importance, fibre-laden turbulent flows have become a subject of significant recent concern~\citep{annurevvoth,chiarini2024104,annurevmarchioli}, prompting numerous relevant computational and experimental studies over the past several decades. A central difficulty is that fibre dynamics couples translation, rotation, inertia, finite-size sampling of the flow and, in wall-bounded configurations, direct or indirect interactions with solid boundaries.

In a wide range of applications, these elongated particles interact with turbulent boundary layers and walls that bound the flow carrying them. These interactions typically lead to complex phenomena like accumulation, deposition, resuspension or even clogging of the flow. Turbulent channel flow constitutes a model laboratory system to study these interactions, providing a simplified yet physically rich framework for examining the interplay between particles, turbulent boundary layer and wall confinement. In fully developed channel turbulence, several regions coexist and exhibit different turbulent features: in the centre, the turbulent fluctuations are nearly homogeneous and isotropic, whereas near the wall, intense mean shear effects dominate, accompanied by strong and anisotropic turbulent fluctuations and coherent structures (streaks, quasi-streamwise vortices, hairpin vortices, etc.)~\citep{pope_2000}. The near-wall velocity scale is the friction velocity $u_\tau=(\tau_{\rm w}/\rho_{\rm f})^{1/2}$, where $\tau_{\rm w}$ is the mean wall shear stress and $\rho_{\rm f}$ the fluid density. It defines the viscous length and time scales,
$\delta_\nu=\nu/u_\tau$ and $\tau_\nu=\nu/u_\tau^2$,
where $\nu$ is the fluid kinematic viscosity.

For rigid fibres, the relevant control parameters depend on which flow scale is used for normalisation. In homogeneous turbulence, the finite-size character of the particle is commonly measured by $\ell/\eta_{\rm K}$, where $\ell$ is the fibre length and $\eta_{\rm K}$ the Kolmogorov length, while its inertia is quantified by a Stokes number $\St=\tau_{\rm p}/\tau_{\rm K}$, with $\tau_{\rm p}$ the particle response time and $\tau_{\rm K}$ the Kolmogorov time. The aspect ratio $\lambda=\ell/d$, where $d$ is the fibre diameter, controls the anisotropy of the hydrodynamic resistance. In wall-bounded turbulence, additional scales enter the problem: the viscous scales $\delta_\nu$ and $\tau_\nu$, and the channel half-height $h$. Physical quantities normalised by viscous units are conventionally denoted by a superscript $+$, yielding in particular $\ell^+=\ell/\delta_\nu$ and $\St^+=\tau_p/\tau_\nu$. The ratio $\ell/h$ measures the degree of confinement by the channel geometry. Thus, $\ell/\eta_{\rm K}$, $\ell^+$ and $\ell/h$ quantify distinct aspects of finite-size effects: sampling relative to dissipative scales, size in wall units, and geometric confinement. When the fibre and fluid densities are not matched, gravity introduces an additional control parameter, for instance the ratio $V_g^+=V_g/u_\tau$ between the quiescent settling velocity~$V_g$ and the friction velocity. This parameter measures the importance of gravitational drift relative to turbulent transport and can strongly affect wall-normal concentration profiles and wall interactions.

Most numerical studies of fibre-laden channel turbulence have focused on the sub-Kolmogorov limit, in which fibres can be represented by point particles endowed with an orientation. In this approach, particles are treated as material points with ellipsoidal geometry, for which the translational and rotational resistance tensors are available analytically~\citep{jeffery,bretherton_1962,brenner19631}. The underlying assumption is that the flow disturbance around the particle is locally well described by a Stokes flow. These numerical works consistently report that short fibres preferentially align with the streamwise direction close to the wall, while their orientation becomes more isotropic towards the channel centre. \citet{zhao_andersson_2016} further showed that, throughout the channel, particles tend to align with the direction of strongest Lagrangian stretching, as previously observed in homogeneous isotropic turbulence~\citep{ni_ouellette_voth_2014,pujara_voth_variano_2019}. This near-wall alignment is nevertheless intermittent: \citet{marchioli2010} reported that, especially at large Stokes number and aspect ratio, fibres remain aligned only for finite times before rotating again. The rotational dynamics was later analysed in detail by \citet{zhao2015,Zhaoetal2019}, who found that elongated particles preferentially spin in the channel centre and tumble more vigorously near the wall. Particle inertia tends to smooth rotational anisotropies in the centre, whereas it enhances them near the wall. The interaction with near-wall turbulent structures also controls preferential sampling and deposition. \citet{morten1,mortensen2008678} showed that ellipsoidal particles tend to segregate into low-speed regions. The deposition and wall-normal flux of fibres were studied by \citet{marchioli2010} and \citet{Yuanetal2018a,Yuanetal2018b}. \citet{marchioli2010} found that translation--rotation coupling modifies the mean wall-normal particle velocity and thereby changes the flux towards the wall, in addition to the direct effect of inertia. \citet{Yuanetal2018a,Yuanetal2018b} showed that the particle flux depends on both inertia and aspect ratio: particles moving towards the wall tend to be associated with sweep events, whereas particles moving away from the wall are associated with ejections. More recently, \citet{Ouchene03102018} studied the acceleration statistics of the particles in the flow while \citet{MichelArcen2021,michel_arcen_2023} focused on the influence of the Reynolds number on the particle dynamics.

Only a few numerical studies have considered finite-size fibres in turbulent channel flow. One possibility is to use slender-body models, in which the fibre diameter is assumed asymptotically small while the fibre length remains finite and can exceed the Kolmogorov scale. Such models are based on slender-body theory~\citep{batchelor_1970,cox_1970,keller_rubinow_1976,anderson}, and again rely on a local Stokes-flow description around each fibre element. The fibre can then be discretised along its length or described by a continuous centreline formulation~\citep{annurevmarchioli}. In turbulent channel flow, slender-body simulations have so far mainly been used to study flexible fibres~\citep{Dottoetal2020,jeremie2021}. A second approach consists of interface-resolved direct numerical simulations~\citep{duo,eshghi,ardekani_brandt_2019,zhang_guo_peng_wang_2025}, in which the flow around the particle surface is explicitly resolved. Such simulations have become increasingly feasible thanks to recent computational advances. In turbulent channel flow, and in contrast with point-particle results, \citet{duo} showed that finite-size fibres increasingly sample high-speed streaks close to the wall as their length increases. Interface-resolved simulations, however, remain computationally expensive and are therefore still limited in terms of statistical convergence and parameter-space exploration.

Experiments naturally access finite-size fibres, since manufacturing, imaging and tracking very small elongated particles remain challenging. As a result, while the numerical literature on turbulent channel flow has long been dominated by point-particle models, several experimental studies over the last decade have investigated fibres whose length is comparable to or larger than the viscous length. A common finding of these studies is that fibres are, on average, preferentially aligned with the streamwise direction near the wall, while their orientations become more broadly distributed towards the centre of the flow. Beyond this robust trend, length, curvature, inertia and settling can substantially modify their translational and rotational statistics.
Using a water table experiment, \citet{Hakanssonetal2013} and \citet{abbasihoseini201513} focused on the interactions between finite-size fibres of different lengths ($\ell^+=7-28$) and the coherent structures close to the wall. They observed and characterised fibre streaks resulting from the accumulation of fibres within near-wall velocity streaks. They found that the velocity of long fibres exhibits a bimodal distribution, whereas shorter fibres preferentially accumulate in low-speed regions. \citet{alipour_depaoli_ghaemi_soldati_2021} investigated the effects of fibre curvature using rigid, nearly inertialess, non-axisymmetric fibres with $\ell^+=11$ and different intrinsic curvatures. These fibres were found to be transported at approximately the fluid velocity in the logarithmic and bulk regions, while lagging the fluid close to the wall. Using the same experimental approach, \citet{alipour_depaoli_soldati_2022} studied the effect of length by varying the flow Reynolds number, and hence the viscous length $\delta_\nu$, obtaining fibres in the range~$\ell^+=6$--$22$. Later, using the same experimental setup and fibres, \citet{guirgui2024} exploited the slight fibre curvature to reconstruct the full rotation rate, including both spinning and tumbling components. These rotation rates, which may be used as proxies for turbulent dissipation~\citep{Zazaetal2026}, were found to increase close to the wall, reach a maximum, and then decrease away from the walls. Their results also showed that the rotation rates of nearly inertialess fibres are dominated by turbulent fluctuations rather than by the mean shear, as further demonstrated by simultaneous measurements of a fibre and its surrounding velocity field~\citep{Giurgiuetal2024b}. \citet{shaik1} and \citet{shaik2023104262} explored the combined effect of fibre length and inertia in a turbulent square duct. They considered fibres with $\ell^+=16$--$67$ and moderate Stokes numbers, $\St^+\leq0.34$. Away from the wall, fibres were observed to lag the fluid, with a lag increasing with both fibre length and inertia. The rotation rates peaked in the buffer layer and remained approximately constant away from the walls. Because the settling velocity was much smaller than the friction velocity, the concentration profile was nearly homogeneous except close to the wall, where a depletion layer was observed, with a thickness decreasing with inertia. No specific direct wall-interaction mechanism was reported.
Finally, \citet{baker_coletti_2022} compared rigid finite-size fibres with $\ell^+=66$ to spheres and disks of comparable inertia, $\St^+=O(10)$, in a turbulent boundary layer. Their particles were thicker than in most previous experiments and had a settling velocity of the order of the friction velocity. As a result, they observed strong settling dynamics, inhomogeneous concentration profiles and significant wall interactions. These effects were accompanied by intense tumbling events close to the wall, with tumbling-rate profiles similar to those reported by \citet{guirgui2024}. Their fibres lagged the fluid in the logarithmic region but preferentially sampled high-speed regions close to the wall.

The picture emerging from these studies is therefore incomplete in two important respects. First, while numerical simulations have mostly focused on point-particle models for sub-Kolmogorov fibres, experiments have explored finite-size effects for fibres whose length exceeds the viscous or Kolmogorov scales. Yet, in the available experimental studies, fibres have remained short compared with the outer length scale of the wall-bounded flow, with $\ell/h$ typically below $0.15$. The regime $\ell/h=O(1)$, in which the fibre length becomes comparable to the channel half-height, remains largely unexplored. In this regime, confinement and wall interactions are expected to be strongly enhanced. At the same time, the fibre length is much larger than the dissipative scales, so that the particle samples velocity differences over inertial-range separations. This suggests dynamics that may differ qualitatively from that of both point-like fibres and shorter finite-size fibres. This expectation is consistent with homogeneous-isotropic-turbulence studies, where long fibres with $\ell/\eta_K\gtrsim20$ display tumbling rates decreasing approximately as $(\ell/\eta_K)^{-4/3}$~\citep{parsa2014,oehmke}. The present work therefore investigates how fibre dynamics is affected when the fibre length is simultaneously much larger than the dissipative scales and comparable to the channel height.
Second, systematic one-to-one comparisons between experiments and simulations performed with matched fibre and flow parameters remain scarce. The second objective of this paper is therefore to perform such a comparison, using direct numerical simulations of the carrier flow coupled to a rigid slender-body model for the fibres. The model builds on our previous simulations of slender flexible fibres in turbulent channel flow~\citep{jeremie2021}. Compared with interface-resolved simulations, this approach is computationally cheaper and allows us to accumulate statistics over a broader range of fibre configurations. At the same time, because the experimental fibres have a finite diameter that is not asymptotically small compared with all relevant flow scales, the comparison provides a stringent test of the slender-body approximation under realistic experimental conditions and helps identify which modelling ingredients should be improved.

This paper is organised as follows. The experimental setup, measurement procedure and fibre properties are described in~\S\ref{sec:setup}. The numerical approach, including the direct numerical simulation of the carrier flow and the rigid slender-body model used for the fibres, is presented in~\S\ref{sec:num}. Results on concentration profiles, translational velocities, orientations and rotation statistics are reported in~\S\ref{sec:results}, together with a systematic comparison between experiments and simulations. We show that fibre length significantly affects the particle dynamics by enhancing confinement and wall interactions, and by modifying the way fibres sample the flow. The agreement between experiments and simulations is generally good for all measured quantities, while the physical origin of the remaining discrepancies is discussed in~\S\ref{sec:discussion}. Conclusions are given in~\S\ref{sec:ccl}.

\section{Experimental setup and procedures}\label{sec:setup}

\subsection{Turbulent water channel platform}

\begin{figure}
  \centerline{\includegraphics[width=1\textwidth]{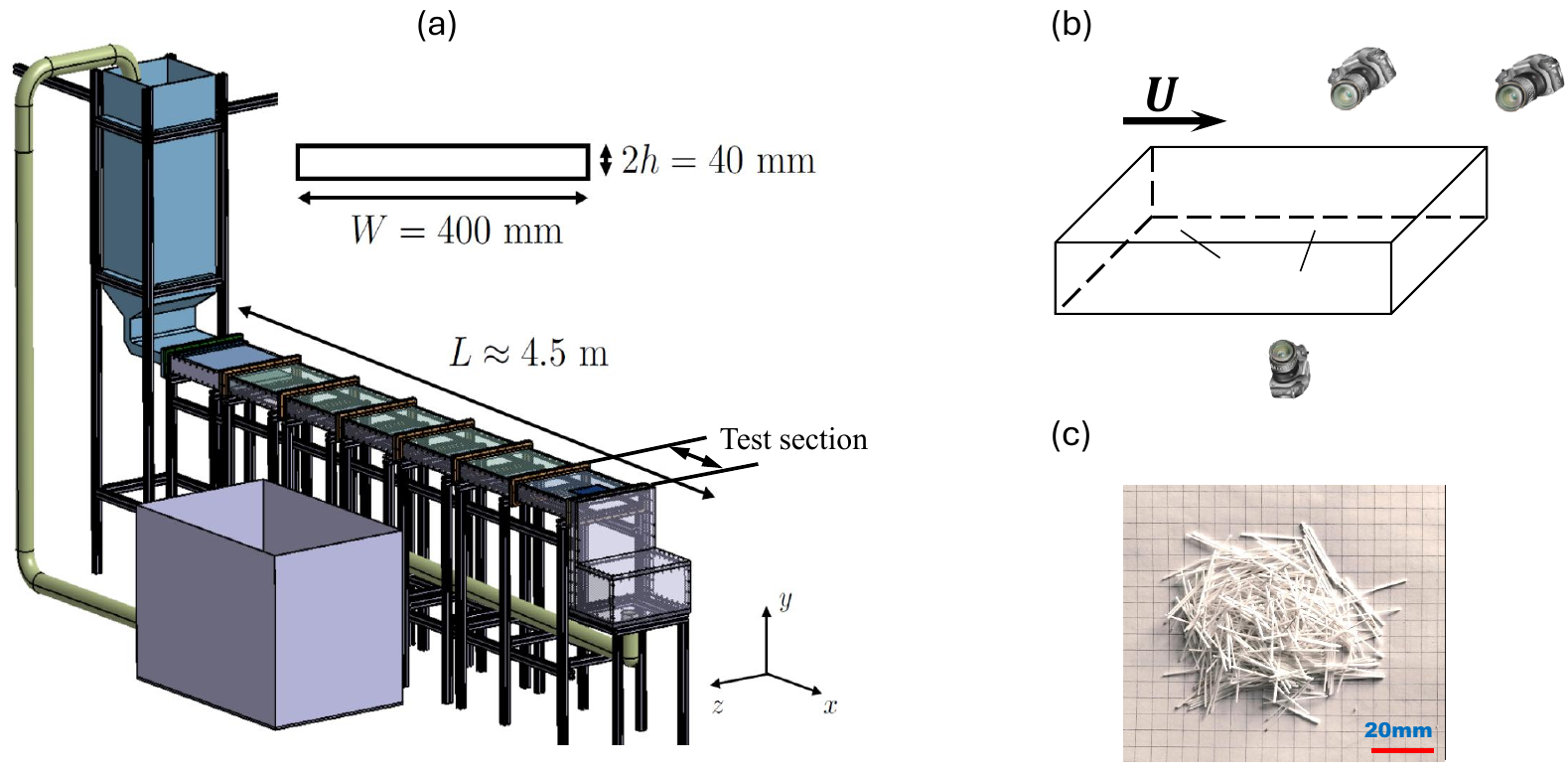}}
  \caption{(a) Sketch of the turbulent water channel platform. (b) Sketch of the camera configuration for the 3D reconstruction of the fibres (see \S\ref{sec:3D}) around the test section. (c) Picture of a pile of the $20$~mm long polystyrene fibres used in this study.}
\label{fig:setup}
\end{figure}

The experimental platform is shown in figure~\ref{fig:setup}(a). It consists of a closed, transparent water channel connected to upstream and downstream reservoirs. Approximately $1~\mathrm{m}^3$ of tap water is circulated by a centrifugal pump whose rotational speed can be adjusted to impose flow rates between $10$ and $50~\mathrm{m}^3\,\mathrm{h}^{-1}$, corresponding to friction Reynolds numbers in the range $200\lesssim\Rey_\tau\lesssim 800$. The flow rate is measured by a magnetic flow meter. When the platform is not in use, water is stored in the downstream reservoir. The channel walls are made of acrylic glass to provide optical access. The channel is $4~\mathrm{m}$ long in the streamwise $x$-direction, with a rectangular inner cross-section of width $W = 400~\mathrm{mm}$ in the spanwise $z$-direction and height $2h = 40~\mathrm{mm}$ in the wall-normal $y$-direction. The channel half-height~$h$ therefore corresponds to $20~\mathrm{mm}$. Note that the wall-normal $y$-direction is parallel to gravity. All measurements were done at the test section, located close to the end of the channel and visible in figure~\ref{fig:setup}(a). The aspect ratio $W/(2h)=10$ of the cross section and the development length upstream of the test section, $175 \,h = 3.5~\mathrm{m}$, were chosen large enough to obtain a fully developed turbulent channel flow with negligible sidewall effects in central spanwise region~\citep{monty2005,stelzenmuller2017},
such as they allow the flow to be statistically homogeneous in the spanwise and streamwise directions~\citep{monty2005,stelzenmuller2017} at the test section.

In all experiments performed in this study, a constant flow rate of $20.4~\mathrm{m}\,\mathrm{h}^{-1}$ has been chosen, corresponding to a bulk velocity of $0.354~\mathrm{m}\,\mathrm{s}^{-1}$ and to a friction Reynolds number $\Rey_\tau =u_\tau h/\nu \approx 386$. The temperature of the water was around $20^\circ\mathrm{C}$, corresponding to a density of $\rho _f\approx 998.5~\mathrm{kg}\,\mathrm{m}^{-3}$ and a kinematic viscosity of $\nu \approx 1\times 10^{-6}\ \mathrm{m}^2\,\mathrm{s}^{-1}$. During the measurements, the temperature slightly increased by less than $0.2^\circ\mathrm{C}\,\mathrm{h}^{-1}$, which represents a limited shift in temperature for measurements typically lasting a few hours.

\subsection{Flow properties}

\begin{figure}
  \centerline{\includegraphics[width=1\textwidth]{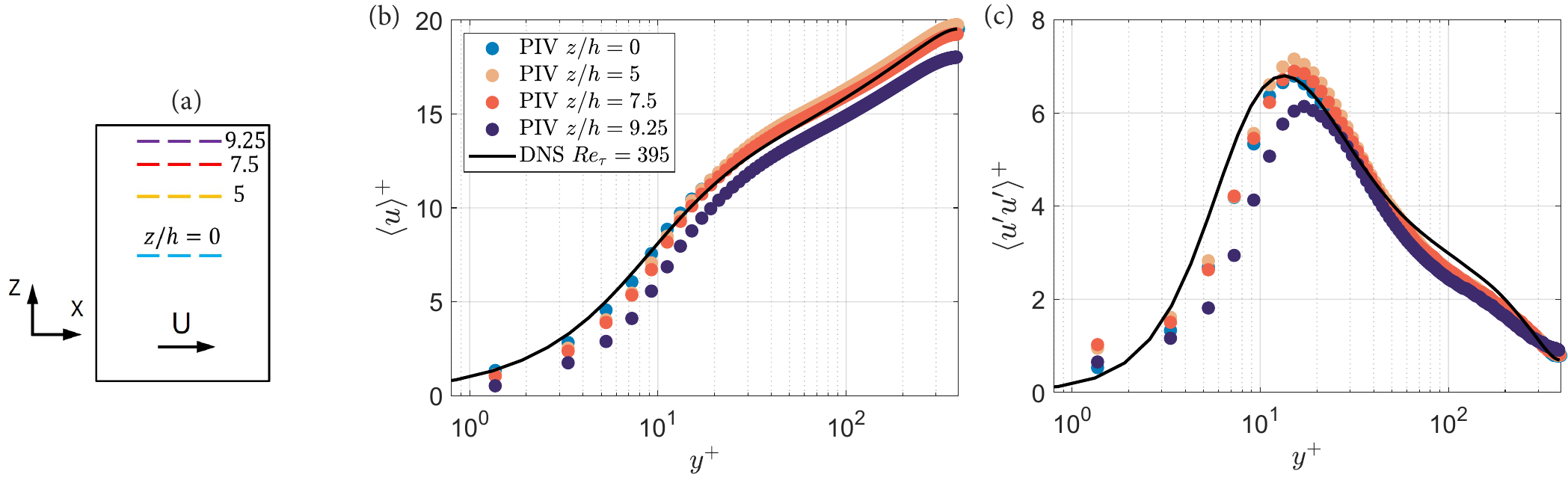}}
  \caption{(a) Spanwise planes where experimental velocity profiles have been measured through PIV. The mid-span plane corresponds to $z/h=0$ while the sidewalls are located at $z/h=\pm 10$. Mean streamwise velocity (b) and Reynolds normal stress (c) profiles in the turbulent channel flow. The points correspond to experimental data at different spanwise locations, while the solid line shows the numerical data.}
\label{fig:sidewall}
\end{figure}

The velocity profiles have been measured through the 2D-2C Particle Image Velocimetry (PIV) technique. A single camera (Phantom TE2010) was placed on the side of the test section, perpendicular to the sidewall. Measurements were carried out in planes parallel to the sidewall (i.e., $x-y$ planes) illuminated using an Nd:YLF laser (Litron LDY301 Lasers; double exposure $10~\mathrm{mJ}/\mathrm{pulse}$ @ $532~\mathrm{nm}$). The laser sheet, with a thickness of about $1~\mathrm{mm}$, was moved to four different spanwise locations in order to evaluate the effect of the sidewalls. The different locations are represented in figure~\ref{fig:sidewall}(a), showing a schematic top view of the test section. The flow was seeded by hollow glass spheres with a mean size of $10~\mu$m and a density of $1100~\mathrm{kg}\,\mathrm{m}^{-3}$, as tracers. Note that no fibres were added in these measurements. Hardware control, data acquisition, and processing were performed with Davis 10.2 (LaVision GmbH). To capture statistically independent velocity fields and to ensure a good convergence of the average quantities, pairs of images have been taken at a frequency of $5~\mathrm{Hz}$ with a time interval between two images of the same pair of $500~\mu\mathrm{s}$. The vector field computation was carried out using a four-pass cross-correlation algorithm with a decreasing interrogation window size from 64 $\times$ 64 to the final size 24 $\times$ 24 pixels with 75$\%$ overlap. In order to take into account the large shear close to the wall, the interrogation windows were allowed to deform close to the wall~\citep{meunierleweke2003}. Additional post-processing and analysis were done through the PIVMat Toolbox~\citep{Moisy2007} using Matlab.

The mean streamwise velocity profile measured at mid-span, $z/h=0$, was used to estimate the friction velocity of the unladen flow, $u_\tau =19.3~\mathrm{mm}\,\mathrm{s}^{-1}$. This gives $\Rey_\tau =u_\tau h/\nu \approx 386$, together with the viscous time and length scales $\tau_{\nu} =\nu /u_\tau ^2=2.7\ \mathrm{ms}$ and $\delta_\nu = \nu/u_\tau = h/\Rey_\tau \approx 52~\mu\mathrm{m}$, respectively. These scales are used hereafter to define wall units, denoted with a superscript $+$. The normalised mean streamwise velocity~$\left \langle u \right \rangle ^+$ and Reynolds normal stress~$\left \langle u'u' \right \rangle ^+$  profiles, measured at different spanwise locations, are shown in figures~\ref{fig:sidewall}(b) and~\ref{fig:sidewall}(c). The profiles obtained for $z/h=0$ (in light blue) are consistent with the numerical simulation profiles (shown by a solid black line) performed at $\Rey_\tau \approx 393$ (see \S\ref{sec:num} for further details on the numerical simulations), thus validating the experimental turbulent flow. Using the turbulent dissipation rate~$\varepsilon^+$ computed in the numerical simulations, we have determined that the Kolmogorov time scale $\tau_\mathrm{K}=(\tau_{\nu}^2/\varepsilon ^+)^{1/2}$ changes from $5.7~\mathrm{ms}$ near the wall to $67.1~\mathrm{ms}$ in the centre of the channel. On the other hand, the Kolmogorov length scale $\eta_\mathrm{K}=\left ( \delta_\nu ^4/\varepsilon ^+ \right ) ^{1/4}$ changes from $75~\mu\mathrm{m}$ near the wall to $259~\mu\mathrm{m}$ in the centre of the channel.

Regarding the velocity and Reynolds normal stress profiles at the different spanwise locations, figure~\ref{fig:sidewall} shows that discrepancies with the mid-span profile at $z/h=0$ are only noticeable in the spanwise plane $z/h=9.25$ closest to the sidewall (located at $z/h=10$). There, the maxima of the mean streamwise velocity $\left \langle u \right \rangle ^+$ and Reynolds normal stress profiles $\left \langle u'u' \right \rangle ^+$ are about $90 \%$ of the maxima of the same profiles in the channel centre. In the 3D reconstruction process later performed for the fibres, only fibres within the range $|z|/h\lesssim4$ are tracked. This ensures that all the fibres considered in the tracking process evolve in a statistically homogeneous flow in the spanwise direction, without feeling the presence of the sidewalls.

\subsection{Fibre properties\label{sec:fib}}

The fibres used in this study are shown in figure~\ref{fig:setup}(c). These fibres are made of polystyrene and have a density of $\rho _p=1050~\mathrm{kg}\,\mathrm{m}^{-3}$, thus being slightly negatively buoyant in water. They are white and opaque, yet appear black in the camera's field of view [figures~\ref{fig:triangula}(a) and~\ref{fig:triangula}(e)], which greatly facilitates observation. Three sets of fibres were investigated in this study, whose properties are summarised in table~\ref{tab:param}. All fibres have the same diameter $d=0.5$~mm ($d^+\approx 10$), and were cut into different lengths $\ell=5,\ 10,\ 20~\mathrm{mm}$ by a home-made device equipped with equally spaced blades. Note that the longest ones have a length equal to the channel half-height~$h$. The aspect ratios $\lambda=\ell/d$ are 10, 20, 40, respectively. For these fibres, the elastic length is estimated as $\ell_{\mathrm{E}}=[cB\tau_{\nu}/(8\pi\rho_\mathrm{f}\nu)]^{1/4}$, where $B$ is the bending modulus, or flexural rigidity, of the fibre and  $c = 2 \log(2\lambda)-1$ is the slender-body shape parameter. First introduced by~\citet{brouzetetal2014}, this length is obtained by balancing the bending relaxation rate of the fibre with the characteristic shear rate~$\tau_{\nu}^{-1}$ of the near-wall flow~\citep{jeremie2021}. In other words, $\ell_{\mathrm{E}}$ represents the critical length above which the fibres start to be flexible. For all fibre lengths considered here, $\ell<\ell_{\mathrm E}\simeq 50~\mathrm{mm}$, so that bending deformations are negligible and the fibres can be treated as rigid bodies.

\begin{table}
  \begin{center}
\def~{\hphantom{0}}
  \begin{tabular}{ccrllccc}
      Set  & ~$d$~[mm]   &   ~$\ell$~[mm]~&  ~$\ell/h$ & ~$\ell^+$&~$St^+$ & $V_\mathrm{g}$~[mm/s] & Color code \\[3pt]
       1  & $0.5$ & $5$~~~ & $0.25$ & ~$96.5$ & ~$15$&~$10.3\pm0.58$ &blue\\
       2   & $0.5$ & $10$~~~ & $0.5$ & ~$193$& ~$19$&~$10.6\pm0.64$ &red \\
       3   & $0.5$ & $20$~~~ & $1$ & ~$386$& ~$24$&~$10.4\pm0.67$ &orange
         \end{tabular}
  \caption{Characteristics of the three experimental fibre sets used in this study.}
  \label{tab:param}
  \end{center}
\end{table}

The relaxation time of a rigid rod is given by (see \S\ref{sec:num} for more details)
\begin{equation}
    \tau_\mathrm{p} = \frac{c \sigma}{8 \pi \mu}=\frac{\rho_\mathrm{p}}{\rho_\mathrm{f}}\frac{c\,d^2}{32\nu},\label{eq:tau_p}
\end{equation}
where $\sigma$ is the fibre linear density and $\mu$ the fluid dynamic viscosity. It is relatively close but not exactly identical to the one for a prolate spheroid given by~\citet{bernstein1994113,annurevvoth}. The Stokes number, as a measure of fibre inertia, can be defined as St$^+=\tau_\mathrm{p}/\tau_{\nu}$ where $\tau_{\nu}=2.7~\mathrm{ms}$ is the viscous time scale. In this study, fibres have a moderate inertia (St$^+ \approx 15$, $19$, and $24$ respectively). Note that the slight changes in Stokes number between the different fibres are only due to changes in aspect ratio, which affect the shape parameter~$c$ in equation~\eqref{eq:tau_p} through a $\log$ function. Experimentally, the most convenient way to change the Stokes number is to modify the fibre diameter~$d$. However, this diameter is fixed in this study, and one can therefore consider the fibre inertia to be almost constant across the three fibre sets used here. In addition, it is also important to note that one can build a Stokes number based on the Kolmogorov time scale~$\tau_\mathrm{K}$, which varies with the wall-normal $y$ distance~\citep{baker_coletti_2022}. This Stokes number decreases approximately from $10$ close to the wall to about $1$ in the centre of the channel for the fibres considered here.

Since the fibres are slightly denser than water, they naturally tend to sediment in the wall-normal direction. Their terminal velocity $V_\mathrm{g}$ has been measured by releasing individual fibres from rest in a tank of quiescent water and recording the time required to reach the bottom. The tank depth, $0.65$~m, was sufficient for the fibres to reach a steady settling velocity. It should be noted that regardless of the initial angle at which particles are added, they reorient quickly to settle with their broad side first, as a result of torques induced by the convective fluid inertia~\citep{Gustavsson_2019}. The measured values of the terminal velocity, reported in table~\ref{tab:param}, are nearly independent of fibre length, with $V_\mathrm{g} \approx 10~\mathrm{mm}\,\mathrm{s}^{-1}$. Indeed, at equilibrium between buoyancy and drag forces, one has
\begin{equation}
    \left ( \rho _p-\rho _f \right ) \frac{\pi d^2 \ell}{4} g\sim \frac{1}{2} C_\mathrm{D}\rho _fV_\mathrm{g}^2d\ell,
\end{equation}
where $g$ is the gravitational acceleration, and $C_\mathrm{D}$ is the drag coefficient of the fibre settling horizontally. This leads to a settling velocity
\begin{equation}
    V_\mathrm{g}=\sqrt{\frac{\left ( {\rho_\mathrm{p}}/{\rho_\mathrm{f}}-1 \right ) \pi d g }{2 C_\mathrm{D}} }, 
\end{equation}
where the fibre length~$\ell$ affects the terminal velocity through the dependence of the drag coefficient~$C_\mathrm{D}$ only. While settling, the fibres have a typical Reynolds number based on their diameter and settling velocity $\Rey_d=V_\mathrm{g} d/\nu \approx5$, leading to a value of $C_\mathrm{D} \approx 4$~\citep{Tritton1959} compatible with the measured values of $V_\mathrm{g}$. 
In the limit of Stokes flow~\citep{cox_1970}, the drag coefficient $C_\mathrm{D}$ scales as:
\begin{equation}
    C_\mathrm{D} \propto \frac{1}{\Rey_d \ln \left ( \ell/d \right ) }.
\end{equation}
For the present conditions, the flow is outside the Stokes regime. By taking into account 
finite~$\Rey_d$ effects in slender body theory, \citet{Joshietal2026} recently claimed that the drag force should be independent of fibre aspect ratio for $\Rey_d \geq 1$, in agreement with our observations. In the following, we therefore consider that the fibres of all three sets settle at the same velocity~$V_\mathrm{g}$. Since this terminal velocity is of the order of $u_\tau/2$, gravitational settling is not negligible. We thus expect it to influence the dynamics of the fibres within the channel, as recently reported by~\citet{baker_coletti_2022}.

Finally, experiments are performed in the dilute regime, in order to make negligible any flow modification due to the presence of the fibres. We typically introduced a few thousand fibres in the full water volume ($\approx 1$~m$^3$), thus leading to a few fibres on average in the measurement volume ($\approx 10^{-3}$~m$^3$), as shown in figure~\ref{fig:triangula}. This typically gives a fibre volume fraction $\phi$ of the order of $10^{-5}$, much smaller than~$1$.
 
\subsection{Fibre detection and reconstruction\label{sec:3D}}

We photograph the fibres using three high-speed cameras (Phantom TE2010) triggered simultaneously, at a frequency of $500$ frames per second. This corresponds to a time interval of $2$~ms between two consecutive frames, which is of the order of the wall time scale~$\tau_{\nu}$ and shorter than the Kolmogorov time scales $\tau_\mathrm{K}$. As shown in figure~\ref{fig:setup}(b), two of them are placed on the side of the test section, inclined at an angle to the sidewall. To be able to focus in the middle of the test section, they are equipped with $150$~mm macro 1:1 lenses (Irix Dragonfly). 
The third camera is placed at the bottom of the test section, perpendicular to the wall, and is equipped with a $40$~mm lens (Nikon). Opposite to the cameras, LED panels are placed in order to observe the particles through backlighting. About $20$ videos of $42000$~images were recorded for each length case. At this frame rate, a video recording typically lasts one and a half minutes and takes several minutes to be transferred to the processing computer.

\begin{figure}
  \centerline{\includegraphics[width=1\textwidth]{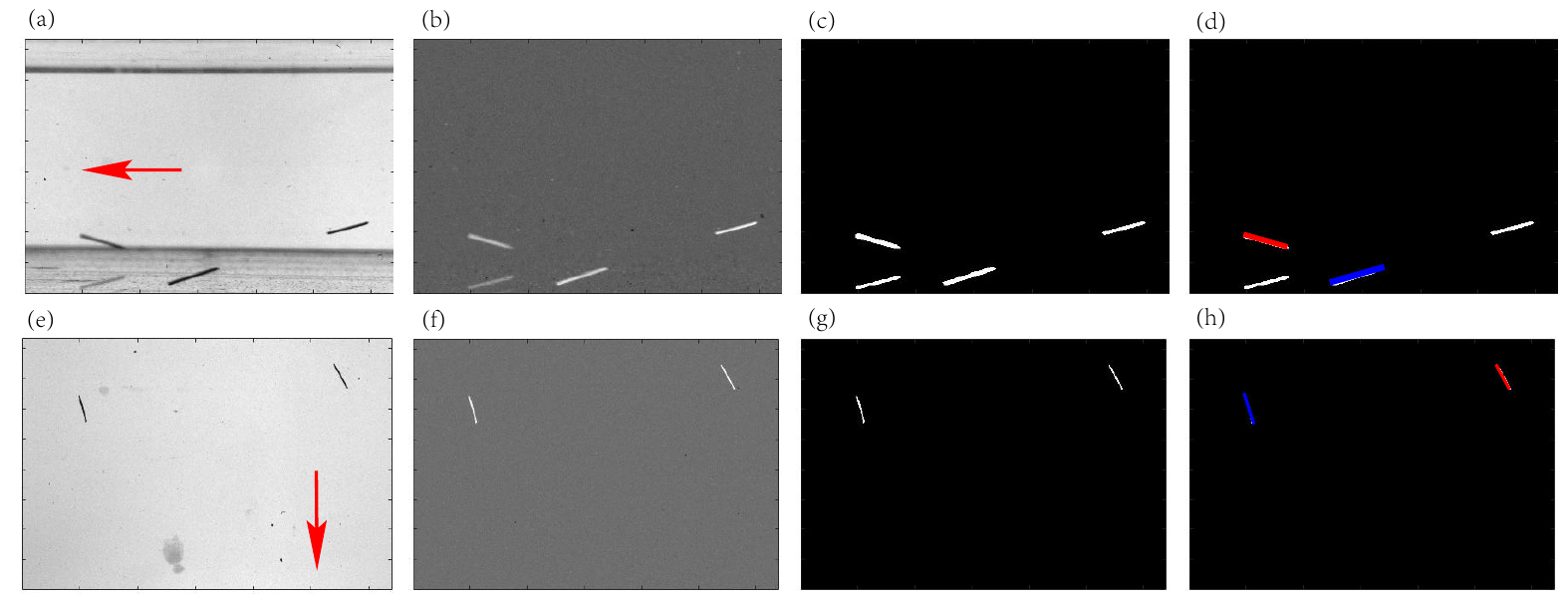}}
  \caption{Examples of fibre detections. The top row shows typical images from a side camera, while the bottom row shows typical images from the bottom camera. Each column exhibits typical images at a given stage of the process: raw images (first column), after background subtraction (second column), after binarisation (third column), and after 3D reconstruction and identification (fourth column). The flow direction is indicated in panels~(a) and (e) by red arrows.}
\label{fig:triangula}
\end{figure}

Examples of raw images containing fibres and recorded simultaneously are shown in figure~\ref{fig:triangula}(a) (from one of two side cameras) and figure~\ref{fig:triangula}(e) (from the bottom camera). Due to parallax effects, horizontally arranged thick black edges are visible in figure~\ref{fig:triangula}(a) and correspond to the top and bottom corners of the channel, with top and bottom walls visible on the image. Such background can be removed by subtracting an image without fibres, as shown in figures~\ref{fig:triangula}(b) and~\ref{fig:triangula}(f). Then, using a threshold on the light intensity, the fibres can be extracted through a binarised image, shown in figures~\ref{fig:triangula}(c) and~\ref{fig:triangula}(g). Note that sometimes we detected a pair of up-down arranged fibres with mirror symmetry along the lower wall (see the lower left corner of figure~\ref{fig:triangula}(a), for example). In this situation, the lower fibre is actually the reflection of the upper fibre on the bottom wall of the channel. This leads to a "ghost" fibre which will be reconstructed outside the channel physical limits. One can therefore remove it afterward in the 3D reconstruction step, by keeping only the fibre voxels whose locations are above the bottom wall position~$y=0$. \
 
Once the fibres are properly detected in each camera, the 3D reconstruction is performed. Cameras are modeled with the pin-hole model~\citep{Gautier2016,oehmke} with $11$ independent parameters characterizing each camera (its position and orientation in the laboratory frame, its focal distances, the coordinates of the projection of the pinhole onto the image, and the skew parameter~\citep{hartley_2004}). The pinhole model relates any point $(x_{3D},y_{3D},z_{3D})$ in the physical space to its projection $(x_{2D},y_{2D})$ onto each camera through the relation $\bq=\mathbb{P}\,\bQ$, where $\bq=[x_{2D},y_{2D},1]^{\!\top}$, $\bQ=[x_{3D},y_{3D},z_{3D},1]^{\!\top}$, and $\mathbb{P}$ is a $3\times4$ projection matrix encoding the $11$ characteristic parameters of cameras. Note that the points in physical and projected spaces $\bQ$ and $\bq$ are expressed using homogeneous coordinates (with a $1$ at the end of the vector), as classically used in computer vision~\citep{hartley_2004}. The matrix $\mathbb{P}$ for each camera is determined by a calibration process where a sphere is moved to a known set of $125$~locations across the measurement volume and imaged by each of the three cameras~\citep{oehmke}. The 3 projection matrices are then optimised altogether, since the 3 cameras have looked together at the same sphere for the same set of coordinates. The sphere was introduced in the test section through a trapdoor located at the end of the test section, on the top wall of the channel, slightly visible in figure~\ref{fig:setup}(a). To account for the refractive index difference between water and air, the calibration process was performed with the channel filled with water. The sub-volume used for fibre detection is around $80 \times 40 \times 150$~mm$^3$. 

To reconstruct fibre position and orientation in three dimensions, we use the shape-from-silhouette or convex-hull method~\citep{Delarosazambranoetal2018,Masuketal2021,Caridi,Ibarraetal2026}. First, the measurement volume is divided it into voxels, with a fine spatial resolution of $0.2$~mm per voxel. This leads to a total of $400 \times 200 \times 750$ voxels. Then, each fibre pixel $\bq=[x_{2D},y_{2D},1]^{\!\top}$ detected on the camera image [figures~\ref{fig:triangula}(c) and~\ref{fig:triangula}(g)] is back-projected to the three-dimensional physical space. However, an image point does not define a unique point in physical space, but a back-projected ray passing through the camera centre. We mark the voxels penetrated by these lines, keeping only those which are at the intersection of the three lines originating from the three cameras. This provides a voxelised visual hull of each fibre, which actually corresponds exactly to its shape~\citep{annurevmarchioli}. Each reconstructed fibre, therefore, corresponds to a list of voxels in the three-dimensional physical space. To validate the reconstruction process, each identified fibre in the physical space is back-projected on the cameras (using the projection matrices) and compared with the binarised images. An example is shown in figures~\ref{fig:triangula}(d) and~\ref{fig:triangula}(h), where fibres of the same colour correspond to the same fibre seen from the three cameras. Fibres which remain white on these images have not been reconstructed properly, either because they are not seen by all the cameras (as it is the case for the fibre on the right of figure~\ref{fig:triangula}(d), which is not visible in figure~\ref{fig:triangula}(h)) or because their centre of mass was located below the known wall position (as it is the case for the fibre on the bottom left of figure~\ref{fig:triangula}(d), which corresponds to the reflection of the red fibre on the bottom wall). In addition, the length tolerance is set at $\ell=\ell_0 \pm5\%$, where $\ell_0$ is the nominal length. Reconstructed fibres that are too long or too short are discarded.

For each reconstructed fibre, we computed the three-dimensional position of the centre of mass~$\bbX=(\bar{x},\bar{y},\bar{z})$, by averaging the coordinates of the fibre voxel points. The three-dimensional fibre orientation unit vector $\bp=(p_x,p_y,p_z)$ is then determined through the eigenvalue decomposition of the inertia tensor~$\mathbbm{I}$~\citep{Caridi}, expressed in the laboratory frame as
\begin{equation}
     \mathbbm{I} \propto \sum_{j=1}^n \begin{bmatrix} (\tilde{y}_j^2+\tilde{z}_j^2) & -\tilde{x}_j\tilde{y}_j & -\tilde{x}_j\tilde{z}_j \\ -\tilde{x}_j\tilde{y}_j & (\tilde{x}_j^2+\tilde{z}_j^2) & -\tilde{y}_j\tilde{z}_j \\ -\tilde{x}_j\tilde{z}_j & -\tilde{y}_j\tilde{z}_j & (\tilde{x}_j^2+\tilde{y}_j^2) \end{bmatrix},
 \end{equation}
where the tilde coordinate~$\tilde{\mathbf{X}}$ represents the original coordinates~$\mathbf{X}$ minus the coordinates of the centre of mass~$\bbX$. The inertia tensor is therefore computed with respect to the fibre centre of mass, by summing over the $n$ voxels constituting the fibre and assuming they represent an equivalent mass in the fibre. After eigenvalue decomposition of~$\mathbbm{I}$, the orientation direction corresponds to the direction of the eigenvector associated with the smallest eigenvalue.

\subsection{Tracking}

\begin{figure}
  \centerline{\includegraphics[width=1\textwidth]{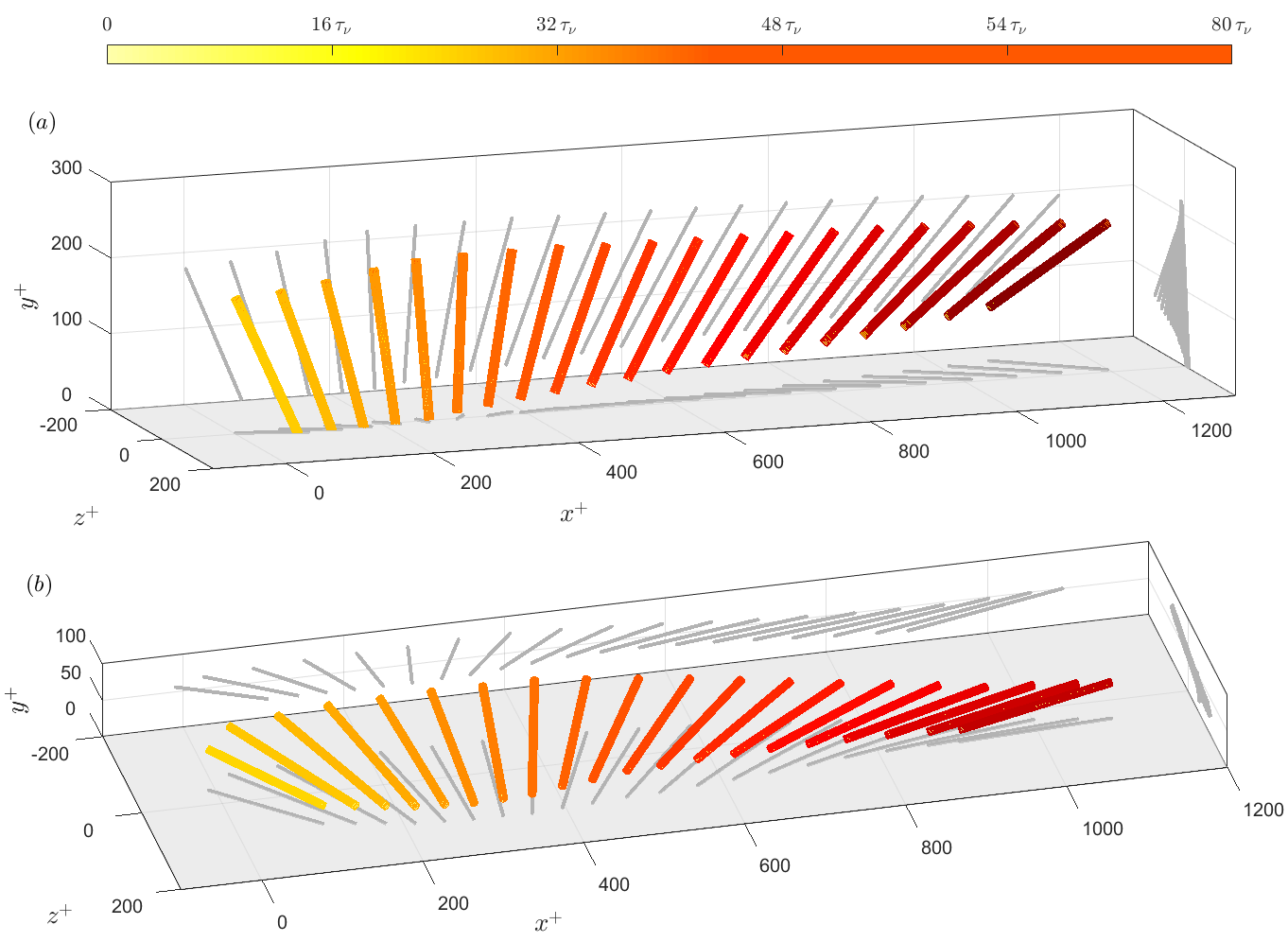}}
  \caption{Typical reconstructed trajectories for $10$~mm long fibres. The colour indicates the time and the flow direction is from left to right. Panel (a) exhibits a pole-vaulting event, where an almost vertical fibre bounces on the bottom wall. Panel (b) shows a kayaking event, where an almost horizontal fibre tumbles in the $x-z$ plane close to the wall. Only one every 5 time steps is displayed for the sake of clarity.}
\label{fig:3dRecon}
\end{figure}

Once each individual fibre has been reconstructed, it is necessary to build trajectories using different images to obtain dynamical quantities, such as fibre velocity and tumbling rate. The fibre tracking process requires the identification of the same fibre in two consecutive frames, using a nearest neighbour algorithm~\citep{ouellette2006,oehmke}. Short trajectories (less than $10$~points) are disregarded to reduce potential statistical errors. On average, a total of $10\times10^3$--$20\times10^3$ trajectories are identified for each fibre length. Two typical reconstructed trajectories are shown in figure~\ref{fig:3dRecon} for $10$~mm long fibres. They illustrate two recurrent types of near-wall motion observed in the experiments: a pole-vaulting event, in which a nearly vertical fibre interacts with and bounces on the bottom wall, and a kayaking event, in which a nearly horizontal fibre undergoes an in-plane tumbling motion close to the wall.

\begin{figure}
  \centerline{\includegraphics[width=.9\textwidth]{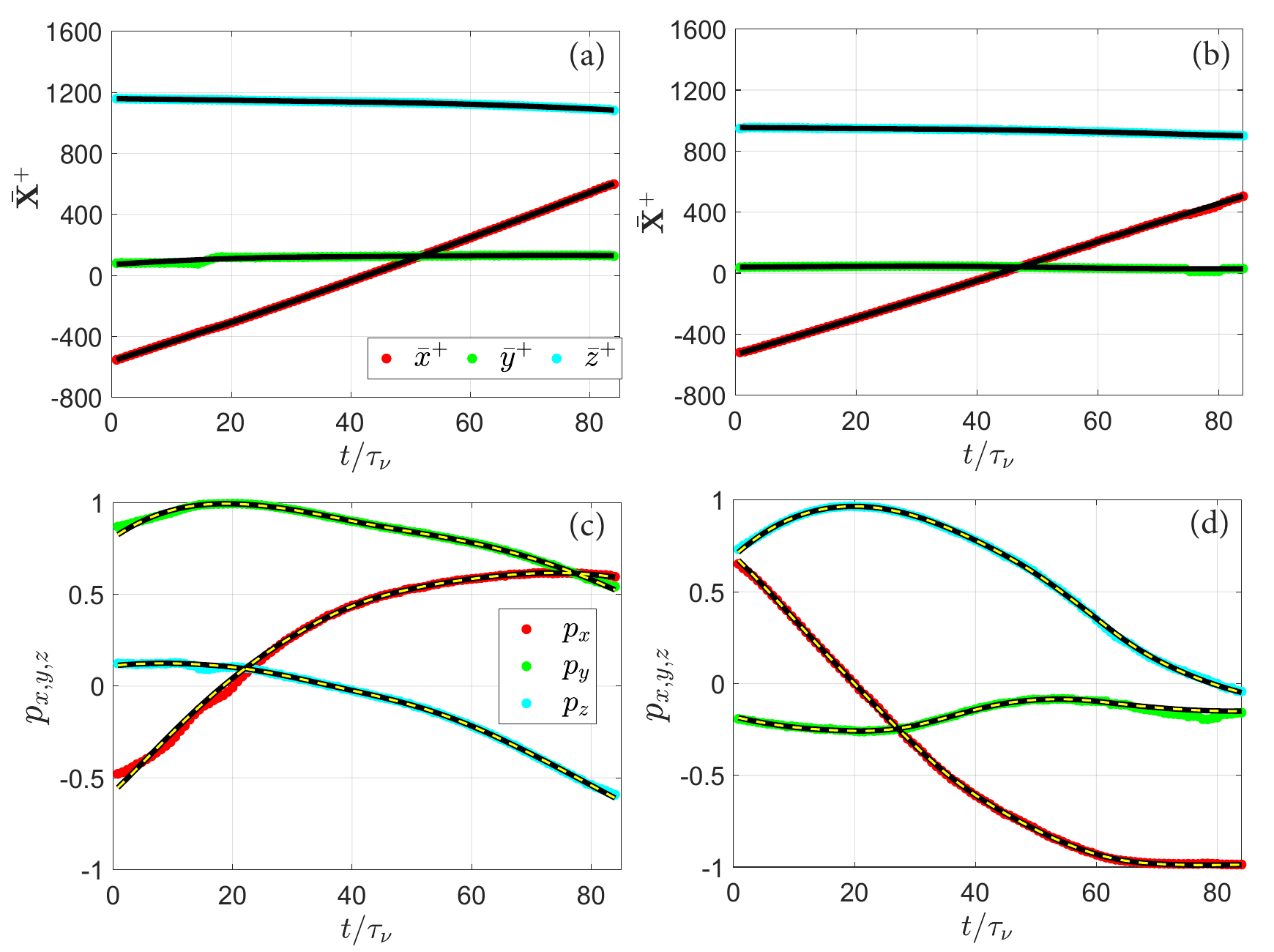}}
  \caption{Time evolution of the centre-of-mass position (top row) and orientation-vector components (bottom row) for the two fibre trajectories shown in figure~\ref{fig:3dRecon}. The first column corresponds to the pole-vaulting event in figure~\ref{fig:3dRecon}(a), while the second column corresponds to the near-wall horizontal tumbling event in figure~\ref{fig:3dRecon}(b). Symbols denote the reconstructed data, black solid lines show the corresponding smoothed signals, and yellow dashed lines in the lower panels show the smoothed orientation vectors after normalization to unit norm. Note that $\bar{\mathbf{X}}^+$ is expressed in the physical laboratory frame, whereas the $\bar{x}^+$ and $\bar{z}^+$ coordinates in figure~\ref{fig:3dRecon} have been shifted for visualization purposes. }
\label{fig:smooth}
\end{figure}

For rigid straight fibres, it is in addition necessary to solve the inherent ambiguity of the direction of $\bp$, which can take two opposite directions along the fibre (positive vs negative). Indeed, this issue becomes severe when computing the tumbling rate $\dot{\bp}$, which relies on the consistency of the orientation direction. A consistency check is performed by flipping the orientation vector~$\bp$ at time~$t$ if $\bp(t)\cdot\bp(t-\textrm{d}t)<0$~\citep{oehmke}.

Once the trajectories have been built and the orientation ambiguity has been solved, the fibre translational velocity and tumbling rate are obtained by differentiating the centre-of-mass position and orientation vector along each trajectory. The time evolution of these quantities is shown in figure~\ref{fig:smooth} for the two example trajectories illustrated in figure~\ref{fig:3dRecon}. Before differentiation, each component was slightly smoothed in order to reduce reconstruction noise. We generally used a robust locally weighted smoothing filter (`rlowess'), as suggested by \citet{Caridi}, with a span parameter equal to $0.4$. For trajectories shorter than $30$~points, a second-order polynomial fit was used instead. The smoothed data are shown by black solid lines in figure~\ref{fig:smooth}.

Because the orientation vector is a unit vector, we also tested the effect of normalising the smoothed orientation vector at each time step before computing its time derivative. The resulting signals are shown by yellow dashed lines in the lower panels of figure~\ref{fig:smooth}. They are almost indistinguishable from the directly smoothed signals, showing that the smoothing procedure does not significantly alter the norm of $\bp$. Nevertheless, this normalisation was retained before differentiating $\bp$, since even small deviations from $|\bp|=1$ may introduce spurious contributions to the tumbling rate.
Time derivatives were computed using second-order centred finite differences in the interior of each trajectory and second-order one-sided finite differences at both ends. However, for the statistics of fibre velocity and tumbling rate, the two endpoints of each trajectory were systematically discarded in order to limit endpoint errors. The tumbling angular velocity, namely the component of the angular velocity associated with changes in the fibre orientation, is then defined as~$\mathbf{\Omega}_\mathrm{T}=\bp \times \dot{\bp}$~\citep{annurevvoth}. Since $|\bp|=1$ and $\bp\cdot\dot{\bp}=0$, the squared tumbling rate satisfies $\left | \dot{\bp }  \right | ^2=\mathbf{\Omega }_{\rm T}\cdot \mathbf{\Omega }_{\rm T}=\Omega_{\mathrm{T},x}^2+\Omega_{\mathrm{T},y}^2+\Omega_{\mathrm{T},z}^2$ .
 
\section{A slender-body model for rigid inertial fibres\label{sec:num}}

The numerical simulations used in this work are designed to reproduce the experimental configuration while retaining a tractable description of the hydrodynamic force acting on long rigid fibres. The carrier flow is a turbulent plane channel flow at friction Reynolds number $\Rey_\tau\simeq 400$. The fibres are modelled as passive, finite-size, inertial slender bodies, whose feedback on the carrier flow, hydrodynamic interactions, and fibre--fibre collisions are neglected. We first introduce the continuous slender-body description, then reduce it to equations for the centre-of-mass position and orientation of a rigid fibre. We finally describe the numerical implementation, the wall-collision model and the simulation parameters.

\subsection{Equations of motion}
The kinematics of long, slender fibres are described by their position $\bX(s,t)\in \mathbb{R}^3$ and velocity $\bV(s,t) = \partial_t\bX(s,t)$ at time $t$, parametrised by the arclength $s\in[-\ell/2,\ell/2]$, with $\ell = \lambda\,d$ denoting the total length of the filament, $d$ the diameter of its circular cross section, and $\lambda$ its aspect ratio. We assume that the filament's motion is governed by the (inertial) slender-body theory~\citep{cox_1970,keller_rubinow_1976}
\begin{equation}
       \sigma\,\partial_t\bV = -\frac{8\pi\mu}{c}\left(\mathbbm{1}-\frac{1}{2}\partial_s\bX\,\partial_s\bX^{\!\top}\right)\left[\bV-\bu(\bX,t)\right] + \boldsymbol{F}_{\rm int}(s,t) + \sigma\,\bg_{\rm eff},
        \label{eq:sbt}
\end{equation}
where $\sigma = m_\mathrm{p}/\ell$ is the fibre's linear mass density, $c = 2\log(2\lambda)-1\approx 2\log\lambda$ is the slender-body shape parameter, the field~$\bu$ denotes the ambient fluid velocity, and $\bg_{\rm eff}=(1-\rho_\mathrm{f}/\rho_\mathrm{p})\bg$ is the buoyancy-corrected gravitational acceleration. The leading-order slender-body drag in equation~\eqref{eq:sbt} assumes that the fibre is slender, $\lambda\gg1$, that the Reynolds number based on the fibre diameter is small, $\Rey_d = |\bV-\bu|\,d/\nu\ll 1$, and that the fluid velocity does not vary significantly over the fibre diameter, namely $d\ll |\bu|/|\nabla\bu|$. End effects and wall-induced hydrodynamic corrections to the drag are also neglected.  The internal forces per unit length, $\boldsymbol{F}_{\rm int}(s,t)$, typically include both tension and elasticity. For instance, in the case of a Cosserat elastic beam \citep[see, \textit{e.g.},][]{Linder}, one has $\boldsymbol{F}_{\rm int} = \partial_s(T\,\partial_s\bX) - B\partial_s^4\bX$, where $T$ is the scalar tension, acting as a Lagrange multiplier enforcing to the inextensibility constraint, and $B$ is the bending modulus (flexural rigidity).

Even without specifying the detailed form of  $\boldsymbol{F}_{\rm int}$, we can infer key properties from the assumption that the filament is passive. In particular, these forces must not contribute to the net linear or angular momenta, implying
\begin{equation} 
	\int_{-\ell/2}^{\ell/2}\boldsymbol{F}_{\rm int}(s,t)\,\dd s = 0 \quad\text{and}\quad \int_{-\ell/2}^{\ell/2} \bX(s,t)\times\boldsymbol{F}_{\rm int}(s,t)\,\dd s = \boldsymbol{0}.
	\label{eq:tot_momenta}
\end{equation}

We now focus on rigid filaments. These are fully described by their centre-of-mass position $\bbX(t) = (1/\ell)\int \bX(s,t)\,\dd s$ and their orientation $\bp(t)$, such that $\bX(s,t) - \bbX(t) = s\,\bp(t)$. An equation for the centre-of-mass motion is obtained by averaging \eqref{eq:sbt} over the filament's arclength. Since internal forces integrate to zero, this yields
\begin{equation}
	\begin{split}
	\frac{\dd\bbX}{\dd t} = \bbV, \qquad \frac{\dd\bbV}{\dd t} = -\frac{1}{\tau_\mathrm{p}} \left(\mathbbm{1}-\frac{1}{2}\bp\bp^{\!\top}\right)\left[\bbV-\bbu \right] + \bg_{\rm eff},\\
    \text{with}\quad \bbu(t) = \frac{1}{\ell} \int_{-\ell/2}^{\ell/2} \bu(\bbX(t)+s\bp(t),t)\, \dd s.
    \end{split}
\label{eq:centre_mass}
\end{equation}
The velocity~$\bbu$ here represents the fluid velocity averaged along the fibre. This dynamics involves a translational response time for the fibre, $\tau_\mathrm{p} = c\,\sigma/(8\pi\mu) = c\,m_\mathrm{p}/(8\pi\mu\ell) \approx (\rho_\mathrm{p}\ell d^2 \log\lambda)/(16\nu\rho_\mathrm{f}\ell)$, as defined in equation~\eqref{eq:tau_p}.

The angular dynamics of the rigid filament is obtained by inserting $\bX(s,t) = \bbX(t)+s\,\bp(t)$ into \eqref{eq:sbt} and subtracting the translational motion \eqref{eq:centre_mass}. This gives
\begin{equation}
	\partial_{t}\bV - \frac{\dd\bbV}{\dd t} = s\,\frac{\dd^2\bp}{\dd t^2} =  -\frac{1}{\tau_\mathrm{p}}  \left(\mathbbm{1}-\frac{1}{2}\bp\bp^{\!\top}\right)\left[s\frac{\dd\bp}{\dd t}-(\bu-\bbu) \right] +\frac{1}{\sigma}\boldsymbol{F}_{\rm int}.
	\nonumber
\end{equation}
Multiplying by $s$ and averaging over arclength, we obtain
\begin{equation}
	\frac{\ell^2}{12}\frac{\dd^2\bp}{\dd t^2} =  -\frac{1}{\tau_\mathrm{p}} \frac{\ell^2}{12}\frac{\dd\bp}{\dd t}  + \frac{1}{\tau_\mathrm{p}\ell}\left(\mathbbm{1}-\frac{1}{2}\bp\bp^{\!\top}\right)\int_{-\ell/2}^{\ell/2} s\,(\bu-\bbu) \dd s +\frac{1}{m_\mathrm{p}} \int_{-\ell/2}^{\ell/2} s\,\boldsymbol{F}_{\rm int}\,\dd s,
	\label{eq:integ}
\end{equation}
where we used that $\bp^{\!\top}(\dd\bp/\dd t)=0$. The constraint on global angular momentum, which reads here $\bp\times\int s\boldsymbol{F}_{\rm int}\,\dd s = 0$, implies that the contribution from internal forces is co-linear to $\bp$, so that there exists $\Gamma(t)$ such that $\int s\boldsymbol{F}_{\rm int}\,\dd s = \Gamma\,\bp$. Projecting equation~\eqref{eq:integ} along $\bp$ then gives
\begin{equation}
	\bp^{\!\top}\frac{\dd^2\bp}{\dd t^2} =  -\left|\frac{\dd\bp}{\dd t}\right|^2 = \frac{6}{\tau_\mathrm{p}\ell^3}\,\bp^{\!\top}\!\int_{-\ell/2}^{\ell/2} s\,(\bu-\bbu)\, \dd s +\frac{12}{m_\mathrm{p}\ell^2} \Gamma.
	\label{eq:integ2}
\end{equation}
Solving for $\Gamma$ and inserting it back into \eqref{eq:integ}, we obtain the final angular equation:
\begin{equation}
	\begin{split}
	\frac{\dd^2\bp}{\dd t^2} &= - \left|\frac{\dd\bp}{\dd t}\right|^2\bp  -\frac{1}{\tau_\mathrm{p}} \left[\frac{\dd\bp}{\dd t}  - \left(\mathbbm{1}-\bp\bp^{\!\top}\right)\bba\right], \\
    &
    \quad\text{with}\quad \bba(t) = \frac{12}{\ell^3}\int_{-\ell/2}^{\ell/2} s\left[\bu(\bbX(t)+s\bp(t),t)-\bbu(t)\right] \dd s.
    \end{split}
	\label{eq:angular_rod_d2p}
\end{equation}
The angular dynamics is therefore governed by the same characteristic time $\tau_\mathrm{p}$ as the translational response. This equality follows from the assumption of a uniform mass distribution along the fibre; a non-uniform distribution would modify the inertia tensor and lead to a different angular response time.

It is useful to introduce the tumbling angular velocity $\boldsymbol{\Omega}_{\rm T}$, defined by
\begin{equation}
\frac{\dd\bp}{\dd t} = \bO_{\rm T}\times\bp, \qquad \bO_{\rm T}\cdot\bp=0.
\label{eq:def_OmegaT}
\end{equation}
This vector contains only the component of the angular velocity perpendicular to the fibre axis; the spin around $\bp$ is irrelevant for the orientation dynamics of an axisymmetric fibre. Taking the cross product of equation~\eqref{eq:angular_rod_d2p} with $\bp$ gives
\begin{equation}
	\frac{\dd\boldsymbol{\Omega}_{\rm T}}{\dd t} = -\frac{1}{\tau_\mathrm{p}} \left[\boldsymbol{\Omega}_{\rm T}-\bp\times\bba\right]\!.
	\label{eq:angular_rod_omeg}
\end{equation}

Some remarks are in order. For small $\tau_\mathrm{p}$, the fibre's translational and angular dynamics reduce, at leading order, to the overdamped equations
\begin{equation}
	\frac{\dd\bbX}{\dd t} = \bbu + \tau_\mathrm{p} \left(\mathbbm{1}+\bp\bp^{\!\top}\right)\bg_{\rm eff}, \quad \frac{\dd\bp}{\dd t} = \left(\mathbbm{1}-\bp\bp^{\!\top}\right)\bba.
\label{eq:overdamped_fibre}
\end{equation}
This is the finite-length slender-body model used, for instance, by \citet{olson1998motion} and \citet{shin2005rotational}. In the additional limit where the fibre length is small compared with the length scale over which the fluid velocity varies, one has $\bba \approx \mathbb{A}\,\bp$ with $\mathbb{A}(t) = \nabla\bu(\bbX(t),t)$, so that
\begin{equation}
	\frac{\dd\bp}{\dd t} = \mathbb{A}\,\bp - [\bp^{\!\top}\mathbb{A}\,\bp]\,\bp,
	\label{eq:jeffery_limit}
\end{equation}
which corresponds to the Jeffery dynamics of an infinitely slender inertialess ellipsoid. The finite-length model can therefore be interpreted as a non-local version of Jeffery's equation, in which the local velocity gradient is replaced by the finite-length directional gradient vector~$\bba$ of the fluid velocity.

We model fibre-wall interactions by a frictionless elastic collision law. When a rigid fibre touches a wall, contact occurs at one of its endpoints, located at $\bbX\pm(\ell/2)\bp$. The velocity of the contact point before impact is $\bV_{\rm c}^- = \bbV^- \pm (\ell/2)\boldsymbol{\Omega}_{\rm T}^-\times\bp$. Let $\bn$ denote the inward unit normal to the wall. If $\bV_{\rm c}^-\cdot\bn<0$, an impulsive correction is applied in the wall-normal direction, leaving tangential velocities unchanged. For a perfectly elastic and frictionless impact, one obtains after the impact
\begin{equation}
	\bbV^+ = \bbV^- + \mathcal{I}\,\bn,\quad \boldsymbol{\Omega}_{\rm T}^+=\boldsymbol{\Omega}_{\rm T}^-\pm\frac{6\,\mathcal{I}}{\ell}\,\bp\times\bn,
	\quad\text{with}\quad \mathcal{I} = -\frac{2\,\bV_{\rm c}^-\cdot\bn}{1+3\left[1-(\bp\cdot\bn)^2\right]}.
\end{equation}

\subsection{Numerical method and simulation parameters}

The carrier flow is computed by direct numerical simulation of the incompressible Navier--Stokes equations in a plane channel. We use the open-source spectral code Channelflow 2.0,\footnote{developed and maintained by the \textit{Emergent Complexity in Physical Systems} (ECPS) team at Swiss Federal Institute of Technology Lausanne (EPFL), \href{https://www.channelflow.ch/}{Channelflow 2.0} (2018).} which employs a fully-dealiased Fourier--Chebyshev--Fourier discretisation in the streamwise, wall-normal and spanwise directions, respectively. Time integration is performed using the third-order semi-implicit backward differentiation formula, with an adaptive time step chosen according to a CFL stability condition. The computational domain has size $L_x\times L_y\times L_z = 4\pi h\times 2h\times 4\pi h/3$, with periodic boundary conditions in the $x$ and $z$ directions and no-slip boundary conditions at the walls. The flow is maintained at constant bulk velocity and reaches a statistically stationary turbulent state with friction Reynolds number $\Rey_\tau=u_\tau h/\nu\simeq 393$. The left panel of table~\ref{tab:param_simu} gives details on the fluid simulation parameters that have been used.

\begin{table}
  \begin{center}
  \begin{minipage}[b]{0.4\linewidth}
  \begin{tabular}{l l}
  	\toprule
       $N_x\times N_y\times N_z$ & $192 \times 193 \times 192$\\
       $L_x\times 2\,h \times L_z$   & $4\,\pi \times 2 \times 4\pi/3$\\
       $U_{\rm b}$ & $0.835$\\
       $\nu$ & $1.25\!\times\!10^{-4}$\\
      $u_\tau$ & $0.049$\\
      $\delta_\nu$ & $0.0025$\\
      $ \tau_\nu$ &  $0.052$ \\ \bottomrule
  \end{tabular}
  \end{minipage}
  \hspace{10pt}
  \begin{minipage}[b]{0.4\linewidth}
  \begin{tabular}{l c c c}
  	\toprule
       $\ell$                    & $0.25$  & $0.5$    & $1.0$\\
       $N_s$                  & $50$     & $100$   & $200$\\
       $\tau_\mathrm{p}$      & $0.8$    & $1.0$    & $1.25$\\
       $\mbox{\it St}^+ = \tau_\mathrm{p}/\tau_\nu$ & $15.4$  & $19.3$  & $24.2$\\
       $N_{\rm f}$          & $10^4$ & $10^4$ & $10^4$\\ \bottomrule
  \end{tabular}
  \end{minipage}
  \caption{Parameters of the numerical simulations, for the fluid (left) and for the fibres (right). Here $U_{\rm b} = \int_{0}^{2h} U(y)\,{\rm d}y/(2h)$ is the bulk velocity, $u_\tau=(\nu |\dd U/\dd y|_{y=0})^{1/2}$ is the friction velocity, $\delta_\nu = \nu/u_\tau$ is the viscous length scale and $\tau_\nu = \nu/u_\tau^2$ is the viscous time scale.}
  \label{tab:param_simu}
  \end{center}
\end{table}

Once the flow has reached a statistically stationary state, passive fibres are introduced at random positions and with random orientations. The dynamics of their centres of mass~\eqref{eq:centre_mass} and of their orientations~\eqref{eq:angular_rod_omeg} are advanced using a second-order Adams--Bashforth scheme with the same time step as the carrier-flow solver. At each time step, the line-averaged velocity~$\bbu$ and directional gradient $\bba$ along each fibre are obtained by trilinear interpolation of the fluid velocity from the Eulerian grid to a set of $N_s$ quadrature points uniformly distributed along the fibre. The line integrals are then approximated using the trapezoidal rule. The right part of table~\ref{tab:param_simu} gives the simulation parameters used for the fibres.

\begin{figure}
    \centering
    \includegraphics[width=1\linewidth]{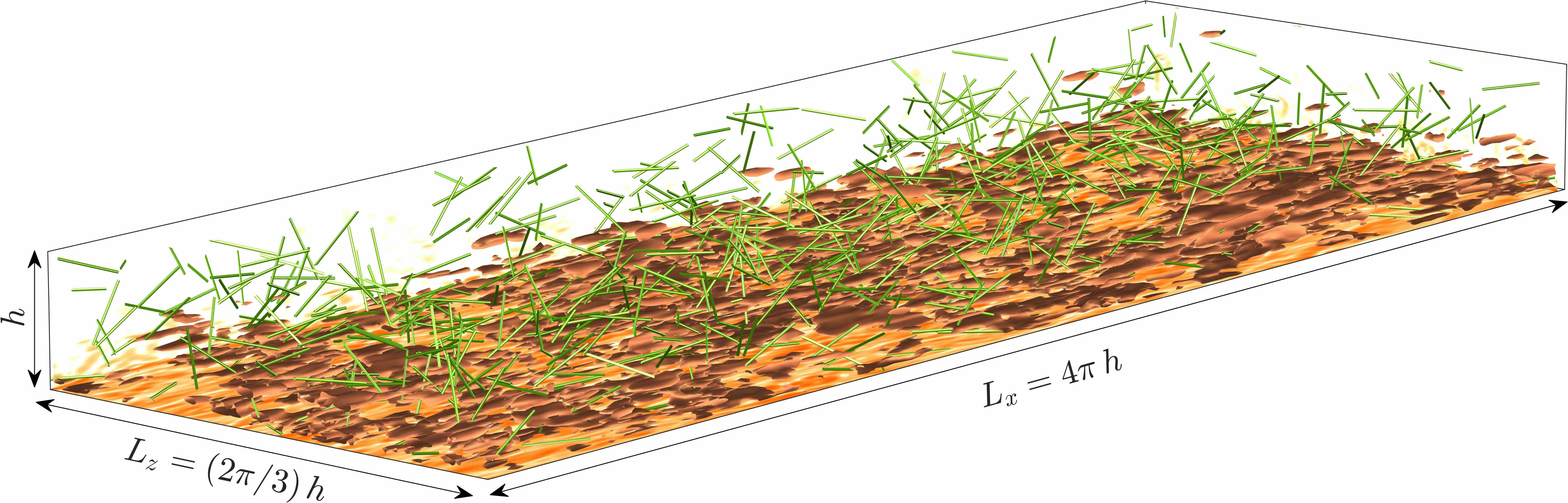}
	\caption{Typical snapshot of a numerical simulation showing the bottom half of the channel, with a volume rendering of the parameter $Q=-(1/2)\,\mathrm{tr}[(\boldsymbol{\nabla}\bu)^2]$ highlighting the rotational structures of the flow. Fibres of length $\ell/h=0.5$ are shown in green.}
    \label{fig:snapshot_num}
\end{figure}

We first performed simulations including the buoyancy-corrected gravity term $\bg_{\rm eff}$ in order to reproduce as closely as possible the experimental conditions. Its magnitude was calibrated so that the terminal settling velocity $V_\mathrm{g}$ of an isolated fibre matched that measured for the experimental fibres discussed in \S\ref{sec:fib}. In these simulations, however, all fibres eventually settled onto the bottom wall, preventing the establishment of a statistically stationary suspension across the channel. This behaviour makes it difficult to accumulate converged wall-normal-conditioned statistics away from the bottom wall. In the following, we therefore focus on simulations without gravity, setting $\bg_{\rm eff}\equiv0$. These simulations should be understood as a reference case for the inertial slender-body dynamics in wall turbulence, in which the effects of finite fibre length, inertia, orientation and wall interactions can be isolated from irreversible sedimentation. A typical instantaneous configuration in this regime is shown in figure~\ref{fig:snapshot_num}. 

\section{Results}\label{sec:results}

\subsection{Concentration profiles}

We first examine the  wall-normal concentration profiles. The channel is divided into $N$ bins along the wall-normal direction~$y$. Let $n_{b}$ denote the total number of fibre observations whose centre of mass lies in bin $b$, with $b = 1,\ldots, N$, and let $N_{f}=\sum_{b=1}^N n_{b}$ be the total number of fibre observations. The normalised concentration at the centre  $y_b^+$ of bin $b$ is defined as
\begin{equation}
    C(y_b^+)=\frac{n_{b}}{N_{f}}\frac{2\Rey_\tau}{\Delta y_b^+}, 
\end{equation}
where $\Delta y_b^+$ is the bin width in wall units. By construction, $\sum_{b=1}^{N}C(y_b^+)\Delta y_b^+/(2\Rey_\tau) = 1$, so that $C=1$ corresponds to a homogeneous distribution across the channel.

\begin{figure}
  \centerline{\includegraphics[width=.8\textwidth]{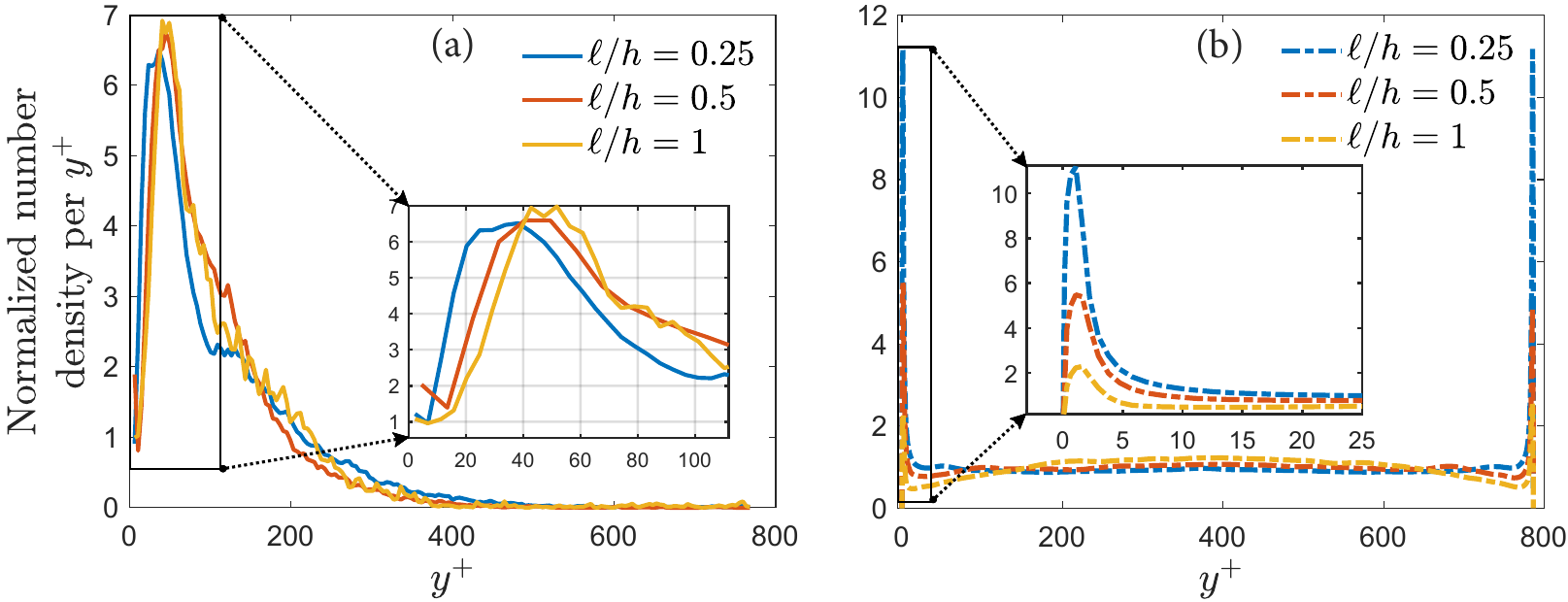}}
  \caption{Experimental (a) and numerical (b) wall-normal concentration profiles for the three fibre lengths. The insets highlight the near-wall concentration maxima.}
\label{fig:density}
\end{figure}

The experimental concentration profiles are shown in figure~\ref{fig:density}(a), with the inset highlighting their near-wall maxima. Starting from the bottom wall, the concentration increases to a pronounced maximum at $y^+ \approx 30$--$50$ and then decreases towards the channel centre. Almost no fibres are observed in the upper half of the channel, above $y^+=400$, demonstrating the strong asymmetry induced by gravitational settling. By contrast, experiments in which the settling velocity is negligible generally report nearly homogeneous concentrations, with only a local enhancement or depletion near the wall \citep{alipour_depaoli_ghaemi_soldati_2021,alipour_depaoli_soldati_2022,shaik1,shaik2023104262}. The dependence on fibre length is weak. The position of the maximum shifts slightly away from the wall as the fibre length increases, but by much less than the factor of four separating the length of the shortest and longest fibres. This modest shift may reflect the stronger geometrical restriction imposed by the wall on long fibres once they depart from streamwise alignment; this effect is examined further in~\S\ref{sec:ori}.

The numerical concentration profiles are shown in figure~\ref{fig:density}(b). Because gravity is absent, they are symmetric with respect to the channel centreline. The concentration remains close to unity through most of the channel but exhibits a sharp maximum at $y^+\approx2$, whose position is essentially independent of fibre length. This near-wall accumulation is consistent with turbophoretic migration driven by fibre inertia \citep{Sardina_Schlatter_Brandt_Picano_Casciola_2012}.

The experimental and numerical concentration profiles therefore differ strongly, primarily because the simulations considered here do not include gravity. To prevent this difference from directly biasing the comparison of kinematic statistics, all subsequent statistics are conditioned on wall distance. When PDFs are reported over finite wall-normal intervals, they are computed in narrow $y^+$ bins and averaged with equal weight, as detailed below.

\subsection{Translational dynamics}

\begin{figure}
  \centerline{\includegraphics[width=.85\textwidth]{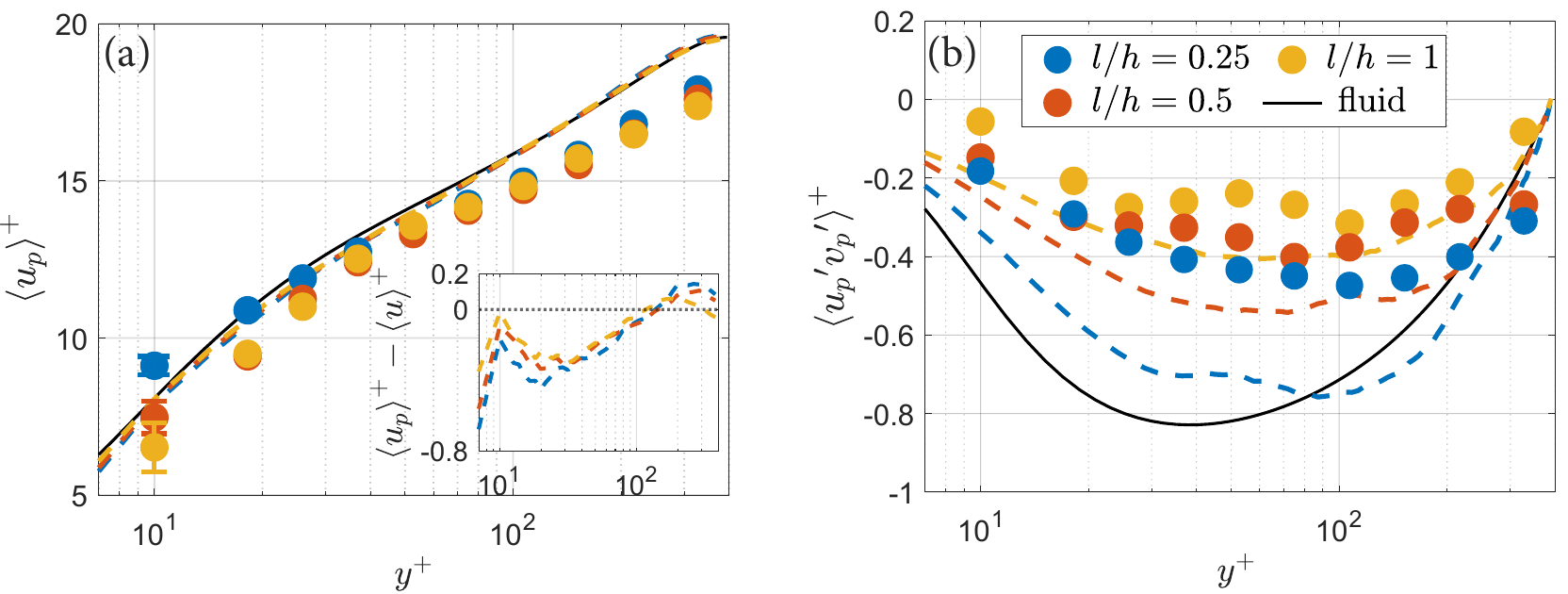}}
  \caption{Mean velocity of the fibre centre of mass in the streamwise $x$-direction (a) and fibre shear Reynolds stress (b) as a function of the wall distance~$y^+$. Experimental data are shown by dots, while numerical data are represented by dashed lines of corresponding colours. The black solid lines show the fluid profile for comparison.}
\label{fig:velocity_mean}
\end{figure}

We now examine the translational statistics of the fibres. Figure~\ref{fig:velocity_mean}(a) shows the conditional mean streamwise velocity of the fibre centre of mass for the experiments (symbols) and simulations (dashed lines). The mean fluid velocity is shown by the black solid line. The corresponding mean wall-normal and spanwise fibre velocities are not shown as they remain statistically indistinguishable from zero throughout the channel. In the simulations, the fibre and fluid velocity profiles remain close. The inset of figure~\ref{fig:velocity_mean}(a) shows a small velocity deficit near the wall and a weak excess in the channel interior, both limited to a fraction of $u_\tau$, with little dependence on fibre length. The experimental velocity differences are substantially larger. In the outer region, $y^+\gtrsim100$, fibres lag the fluid by an amount of order $u_\tau$, with no systematic dependence on length. Close to the wall, however, the mean fibre velocity decreases monotonically with length: the shortest fibres move faster than the local mean fluid, whereas the longest fibres remain slower.

Previous experiments provide a mixed picture of the streamwise velocity deficit. A fibre lag has been reported by \citet{shaik1} and~\citet{shaik2023104262} in the channel centre, and by \citet{baker_coletti_2022} in the logarithmic layer. \citet{shaik1} and~\citet{shaik2023104262} additionally observed a larger deficit for longer fibres, which they attributed to increased drag. No comparable length dependence is observed here, even though the present fibres are longer both in wall units and relative to the channel size. Moreover, the settling measurements in quiescent water indicate no dependence of the drag coefficient on fibre length. By contrast, \citet{capone2017189} and \citet{alipour_depaoli_ghaemi_soldati_2021,alipour_depaoli_soldati_2022} found that fibres are transported at approximately the local mean streamwise fluid velocity away from the wall.

Figure~\ref{fig:velocity_mean}(b) provides complementary information through the fibre shear Reynolds stress. As for the fluid, this quantity is negative for the fibres in both experiments and simulations, indicating that positive wall-normal velocity fluctuations are, on average, correlated with negative streamwise velocity fluctuations. Its magnitude is nevertheless smaller in the experiments than in the simulations over most of the channel. Thus, although fibres moving away from the wall generally have a lower streamwise velocity, this instantaneous correlation is weaker experimentally. In particular, the larger experimental mean streamwise velocity deficit is not accompanied by a stronger ejection-type covariance. The relation between this result, gravitational settling, and the sampling of near-wall velocity structures is discussed in~\S\ref{sec:discussion_gravity}.

Near the wall, the excess velocity measured for the shortest fibres is consistent with the preferential sampling of high-speed regions reported in some previous experiments, notably by \citet{baker_coletti_2022}. Other studies, however, have
reported weaker excess velocities or even a fibre lag, indicating that the near-wall trend depends on fibre length, inertia and flow configuration \citep{abbasihoseini201513,capone2017189,alipour_depaoli_ghaemi_soldati_2021,alipour_depaoli_soldati_2022,shaik1,shaik2023104262}. In the present experiments, the near-wall velocity decreases clearly with fibre length. This behaviour is consistent with the shortest and longest fibres preferentially sampling high- and low-speed regions, respectively, as examined below using the velocity PDFs. Unlike the experiments, the simulations do not show a near-wall excess velocity for the shortest fibres. This discrepancy is discussed in \S\ref{sec:discussion}.

\begin{figure}
  \centerline{\includegraphics[width=.85\textwidth]{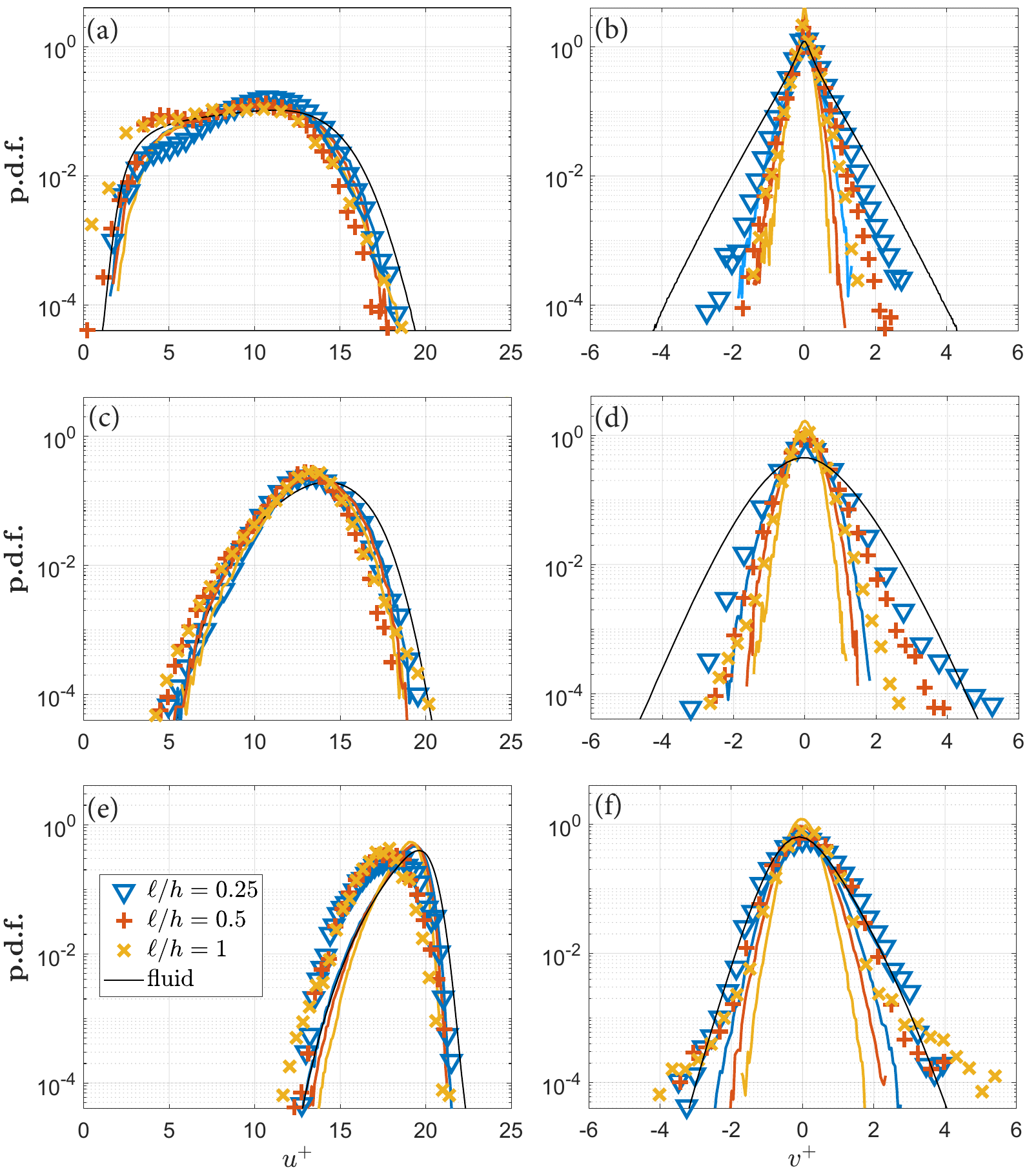}}
  \caption{Probability density functions of the streamwise and wall-normal components of the fibre centre of mass velocity. Top row corresponds to particles located  in the buffer layer ($5 \leq y^+\leq 30$), middle row to particles in the logarithmic region ($30 \leq y^+\leq 60$), while bottom row corresponds to particles located within the bulk of the flow ($200 \leq y^+ \leq 400$). Experimental data are shown by points, while numerical data are represented by dashed lines of corresponding colours. The fluid data are shown by a black line in each panel for comparison.}
\label{fig:vpdfexp}
\end{figure}

We further examine the fibre velocities through probability density functions (PDFs) of their streamwise and wall-normal components in three representative regions of the flow: the buffer layer ($5 \leq y^+ \leq 30$), the logarithmic layer ($30 \leq y^+ \leq 60$), and the channel bulk ($200 \leq y^+ \leq 400$). To prevent the inhomogeneous concentration profiles from biasing these distributions, a conditional PDF is first computed in successive bins of width $\Delta y_b^+ = 5$. The PDFs are then averaged with equal weight over each of the broader wall-normal intervals. The same procedure is applied to the experimental, numerical, and fluid data.

The velocity PDFs are shown in figure~\ref{fig:vpdfexp}. The streamwise distributions confirm the trend observed in figure~\ref{fig:velocity_mean}(a): length dependence is weak, except near the wall, and fibres lag the fluid in the experimental channel bulk but not in the simulations. In the buffer layer [figure~\ref{fig:vpdfexp}(a)], the fibre velocity distributions are bimodal and differ markedly from the fluid distribution. The shortest fibres exhibit a greater probability of high-speed events, whereas the longest fibres exhibit more low-speed events. This is consistent with the shortest and longest fibres preferentially sampling high- and low-speed near-wall regions, respectively, thereby accounting for their opposite mean velocity differences.

For the wall-normal component, the fluid distributions are generally broader than the fibre distributions. An exception occurs in the channel interior [figure~\ref{fig:vpdfexp}(f)], where the experimental fibre and fluid PDFs nearly coincide. In both experiments and simulations, the width of the fibre distribution decreases with increasing length in all three regions. This trend is consistent with the stronger spatial averaging of turbulent velocity fluctuations by longer fibres, which suppresses extreme velocity events. Pronounced positive tails are observed for the shortest fibres ($\ell/h=0.25$) in the logarithmic layer [figure~\ref{fig:vpdfexp}(d)] and for the longest fibres ($\ell/h=1$) in the channel interior [figure~\ref{fig:vpdfexp}(f)]. These tails are associated with pole-vaulting events, such as that shown in figure~\ref{fig:3dRecon}(a), during which wall contact produces a large positive wall-normal velocity. A wall-contacting fibre can attain an almost wall-normal orientation when its centre lies near $y^+=\ell^+/2$, corresponding approximately to $y^+=50$ and $200$ for the shortest and longest fibres, respectively. The pole-vaulting signatures therefore occur in different wall-normal intervals for the different lengths. The experimental wall-normal distributions are also slightly skewed towards positive velocities. The possible role of gravitational settling in producing this asymmetry is discussed in \S\ref{sec:discussion}.

\subsection{Orientation statistics\label{sec:ori}}

\begin{figure}
  \centerline{\includegraphics[width=1\textwidth]{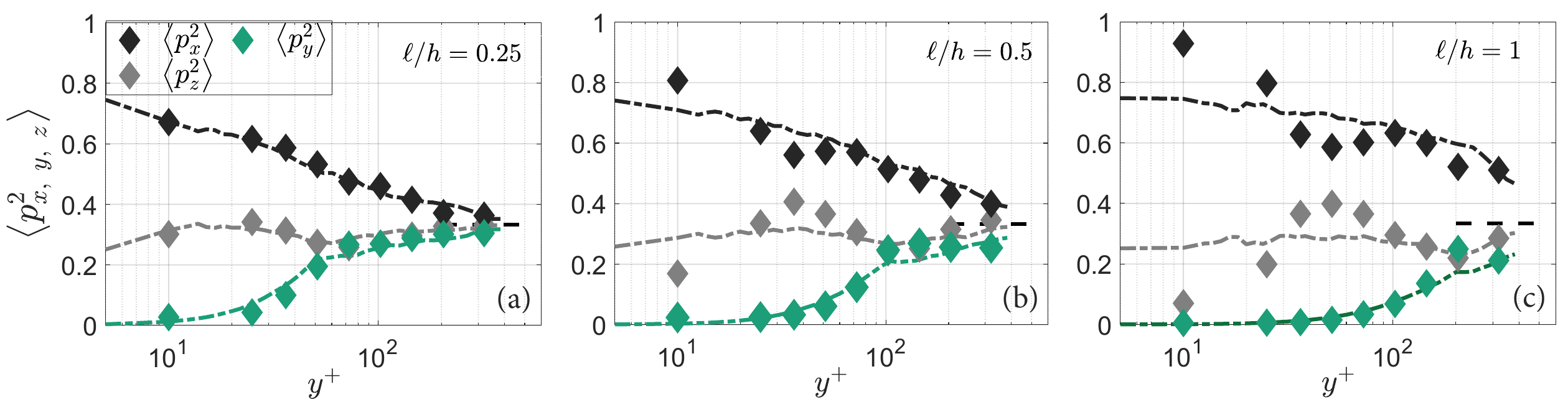}}
  \caption{Mean-square values of the different components of the orientation unit vector~$\boldsymbol{p}$ as a function of the wall distance~$y^+$. The different panels correspond to different lengths: (a)~$\ell/h=0.25$, (b)~$\ell/h=0.5$ and (c)~$\ell/h=1$. Experiments are represented using diamonds while numerical simulations are shown using dashed dotted lines. The horizontal dashed lines at $\langle p_i^2 \rangle = 1/3$ for large wall distances indicate isotropic orientation.}
\label{fig:Orientation}
\end{figure}

We next examine the orientation statistics. Figure~\ref{fig:Orientation} shows the conditional mean-square components of the orientation unit vector $\boldsymbol{p}=(p_x,p_y,p_z)$ for the experiments (diamonds) and simulations (dash-dotted lines), with one panel for each fibre length.

Consistently with previous experimental and numerical studies of shorter fibres in turbulent channel flow~\citep{annurevvoth,alipour_depaoli_ghaemi_soldati_2021,shaik1,baker_coletti_2022}, the fibres align preferentially with the streamwise direction $x$ near the wall. As the centre of mass moves away from the wall, the normal component $\langle p_y^2 \rangle$ increases and the orientation becomes progressively more isotropic (represented as a horizontal dashed line at $1/3$). The longest fibres nevertheless retain appreciable anisotropy in the channel bulk. This residual anisotropy may result from their length being comparable to the channel half height~$h$, so that a single fibre samples substantially different wall-normal regions even when its centre lies near the channel centre. 

The agreement between experiments and simulations is good for the shortest fibres, $\ell/h=0.25$, but systematic differences appear for the two longest fibre sets, mainly in $\langle p_x^2\rangle$ and $\langle p_z^2\rangle$. Immediately adjacent to the wall, the experimental fibres with $\ell/h=0.5$ and $1$ are more strongly aligned with the streamwise direction than their numerical counterparts. At intermediate wall distances, the trend reverses: over approximately $30\lesssim y^+\lesssim90$, the experimental fibres show weaker streamwise and stronger spanwise alignment than the simulations. The interval is narrower for $\ell/h=0.5$, approximately $30\lesssim y^+\lesssim50$, and broadens for $\ell/h=1$. In this region, the experimental values approach $\langle p_x^2\rangle\simeq0.6$, $\langle p_z^2\rangle\simeq0.4$ and $\langle p_y^2\rangle\simeq0$. The fibres are therefore almost entirely confined to the wall-parallel $x$--$z$ plane, while retaining only a moderate preference for the streamwise direction. The discrepancy increases systematically with fibre length, and is largest for the longest fibres. Its possible origin is examined in \S\ref{sec:discussion}.

\begin{figure}
  \centerline{\includegraphics[width=1.\textwidth]{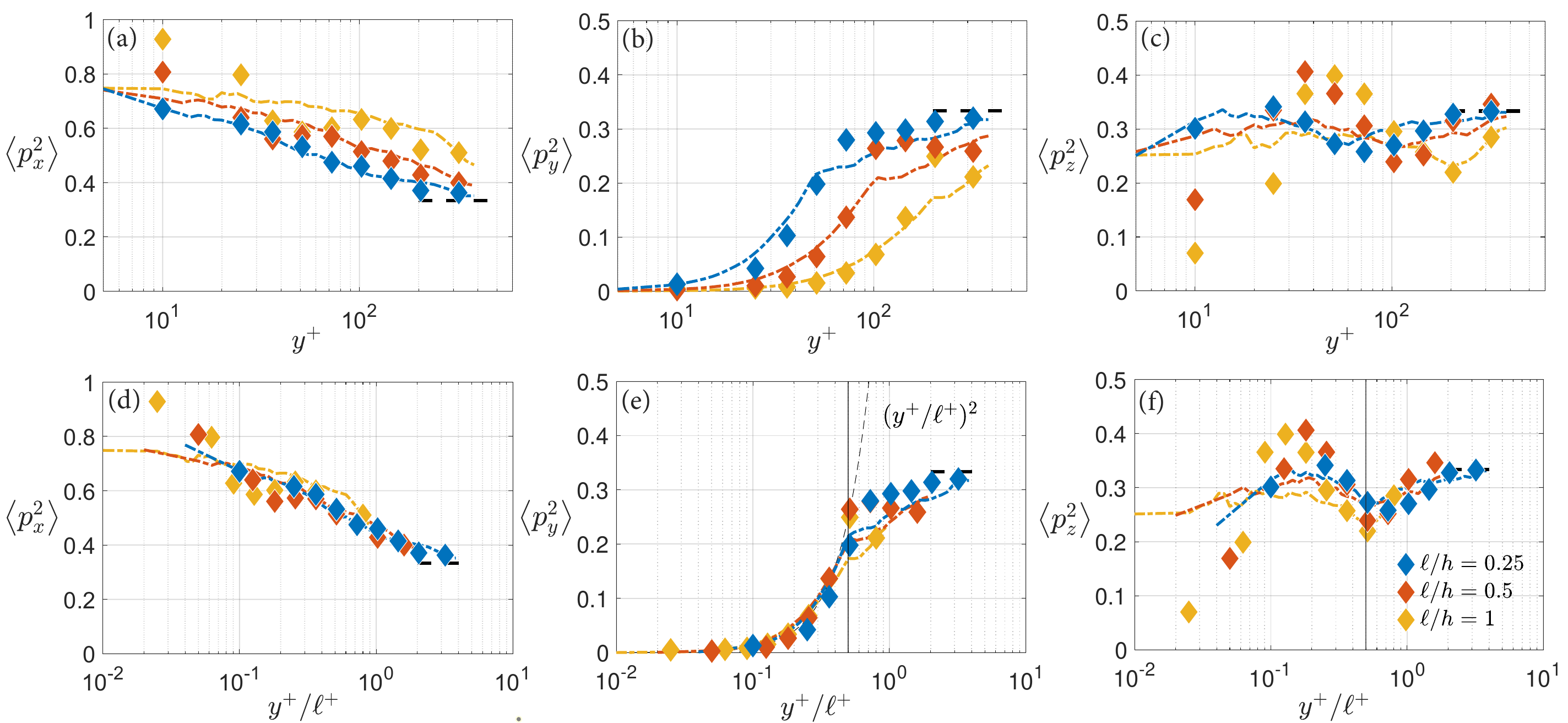}}
  \caption{Mean-square values of the different components of the orientation vector~$\boldsymbol{p}$ as a function of the wall-normal distance of the fibre centre from the wall~$y^+$ (top row) or of the wall distance normalised by fibre length~$y^+/\ell^+$ (bottom row). The data are the same as in figure~\ref{fig:Orientation}, but the different panels now correspond to different components: $\langle p_x^2 \rangle$ (a) and (d), $\langle p_y^2 \rangle$ (b) and (e), $\langle p_z^2 \rangle$ (c) and (f). Experiments are represented using diamonds, while numerical simulations are shown using dashed dotted lines. The horizontal dashed lines at $\langle p_i^2 \rangle = 1/3$ for large wall distances indicate isotropic orientation. }
\label{fig:NormOrien}
\end{figure}

To isolate the dependence on fibre length, the data from figure~\ref{fig:Orientation} are reorganised in figure~\ref{fig:NormOrien}, with one panel for each orientation component. The top row uses the wall distance~$y^+$, whereas the bottom row uses the rescaled distance~$y^+/\ell^+$. This rescaling produces an approximate collapse of most of the experimental and numerical data, showing that the fibre length controls the near-wall evolution of the orientation statistics. The clearest collapse is obtained for $\langle p_y^2\rangle$. When $y^+<\ell^+/2$, non-penetration of the wall imposes the geometrical constraint $y^+-(\ell^+/2)|p_y|\geq0$, or equivalently $|p_y|\leq 2y^+/\ell^+$. The admissible range of wall-normal orientations therefore grows linearly with $y^+/\ell^+$. The observed scaling $\langle p_y^2\rangle\propto (y^+/\ell^+)^2$ shows that the statistics are controlled by this geometrical bound. At $y^+=\ell^+/2$, all wall-normal orientations become geometrically accessible, which explains the change in slope visible in figure~\ref{fig:NormOrien}(e). The streamwise and spanwise components also collapse approximately when plotted against $y^+/\ell^+$, except over the interval in which the experimental and numerical orientation statistics differ.

\begin{figure}
        \centerline{\includegraphics[width=1\textwidth]{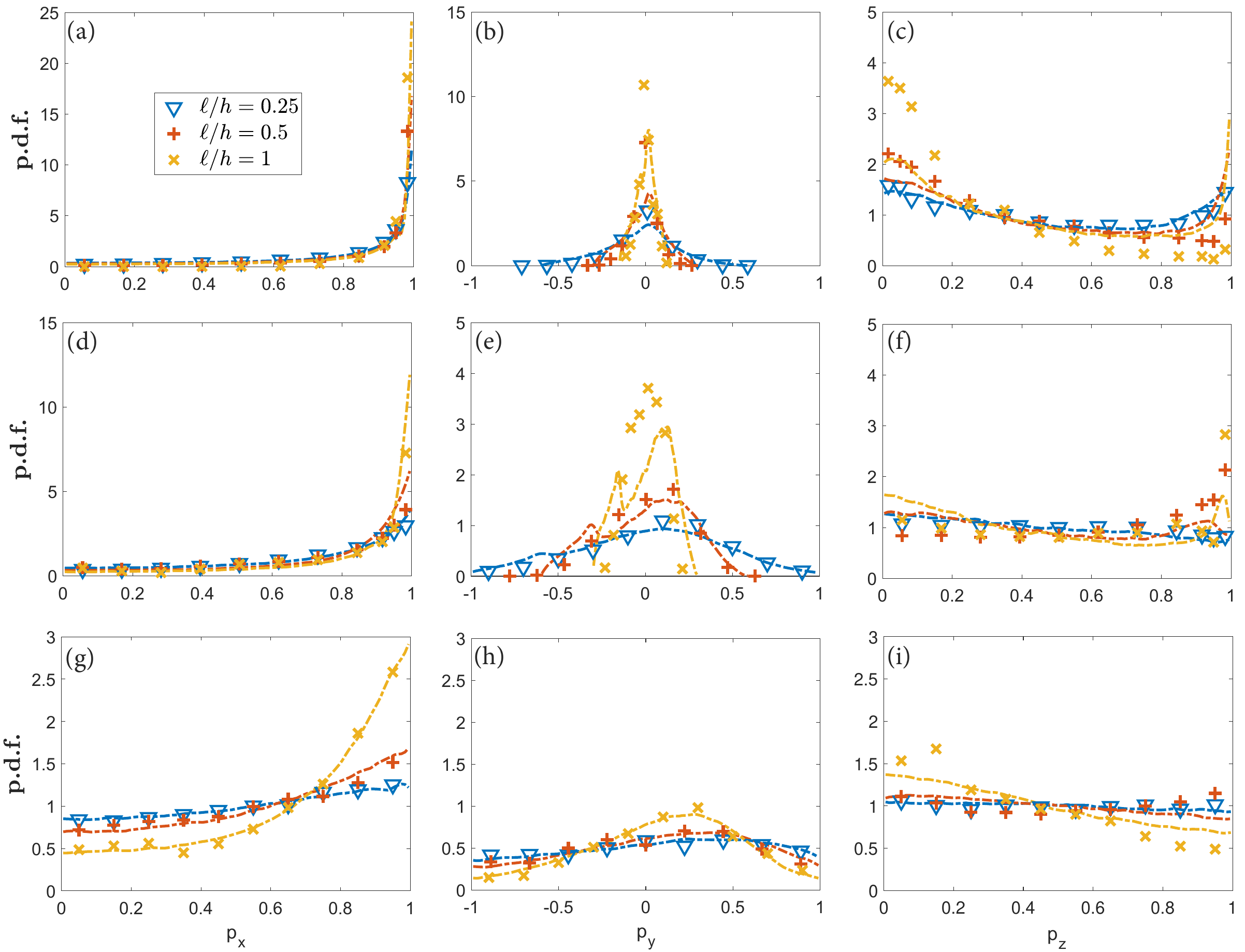}}
  \caption{Probability density functions of the orientation vector components for three different regions of the flow: $y^+ \in [5,30]$ (top row), $y^+ \in [30,60]$ (middle row) and $y^+ \in [200,400]$ (bottom row). Experiments are represented by symbols and simulations by lines. Both are coloured as a function of the fibre length. The different components are shown in different columns.}
\label{fig:OrienPDFall}
\end{figure}

Figure~\ref{fig:OrienPDFall} shows the PDFs of the three orientation components in the same wall-normal intervals used for the velocity statistics in figure~\ref{fig:vpdfexp}. Because the fibres are head-tail symmetric, all orientation vectors are mapped onto the hemisphere $p_x>0$. Under this convention, $p_y>0$ corresponds to a nose-up configuration, in which the leading end lies above the trailing end, whereas $p_y<0$ corresponds to a nose-down configuration. As already reported by \citet{baker_coletti_2022}, the $p_y$ distributions are weakly asymmetric towards positive values, indicating a slight preferential nose-up orientation for all three lengths.  

For the shortest fibres, experiments and numerical simulations agree well for all components and wall-normal regions. For the longer fibres, the PDFs confirm the differences already identified from the second moments. In the buffer layer, fibres are strongly aligned with the streamwise direction, with the alignment becoming sharper as the length increases. In the logarithmic layer, the longer experimental fibres are less aligned with the streamwise direction and more oriented in the spanwise direction than their numerical counterparts. The simulations also show a bimodal distribution of $p_y$ for the longer fibres, whereas no clear bimodality is observed experimentally. The possible influence of fibre--wall contact dynamics is discussed in \S\ref{sec:discussion}. In the bulk of the flow, the experimental and numerical $p_x$ and $p_y$ distributions agree reasonably well, whereas some differences remain for $p_z$. 

\subsection{Tumbling dynamics}

\begin{figure}
  \centerline{\includegraphics[width=.68\textwidth]{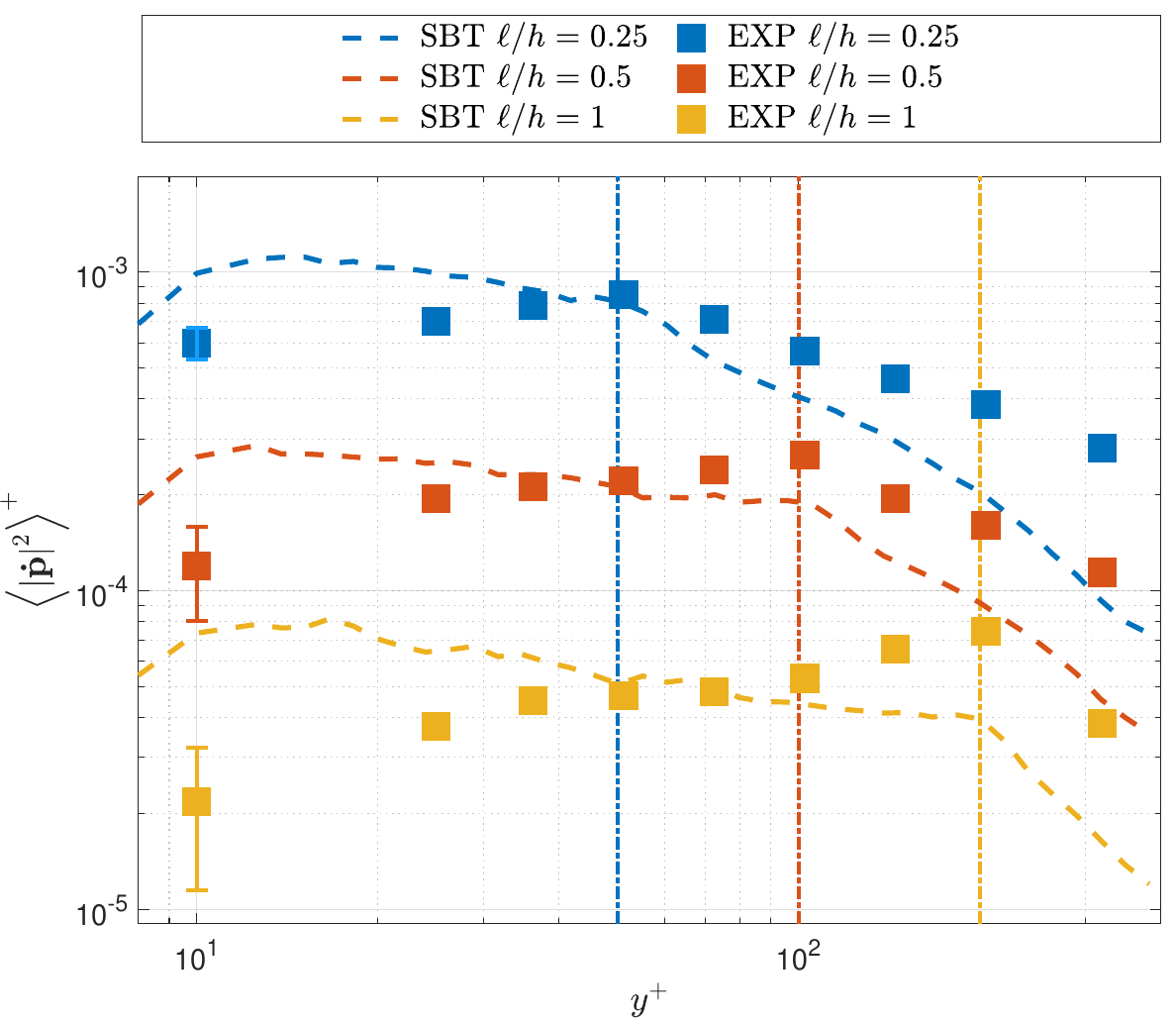}}
  \caption{Tumbling rate as a function of the wall distance. Experimental data are represented by squares, while numerical data are shown using dashed lines. The different colours indicate different fibre lengths. The vertical dashed dotted lines indicate $y^+=\ell^+/2$.}
\label{fig:tumbrate}
\end{figure}

We finally examine the tumbling dynamics. Figure~\ref{fig:tumbrate} shows the conditional mean-square tumbling rate $\langle|\dot{\boldsymbol{p}}|^2\rangle=\langle|\boldsymbol{\Omega}_{\rm T}|^2\rangle$ for the three fibre lengths. Overall, the experiments and simulations agree reasonably well. Close to the wall, the tumbling rate is approximately constant, with a weak increase in the experiments and a weak decrease in the simulations. Beyond $y^+=\ell^+/2$, marked by the vertical dash-dotted lines, it decreases with wall distance. This decrease is consistent with the weakening of the characteristic turbulent velocity gradients, and of the turbulent dissipation rate, away from the wall~\citep{Zazaetal2026}. The experimental profiles exhibit a local maximum near $y^+=\ell^+/2$. The numerical profiles do not show an equally distinct maximum, but display a change of slope at approximately the same location. Above $\ell^+/2$, the numerical tumbling rate decreases more rapidly than its experimental counterpart. A possible connection with settling-induced sampling is discussed in \S\ref{sec:discussion}.

\begin{figure}
  \centerline{\includegraphics[width=1.\textwidth]{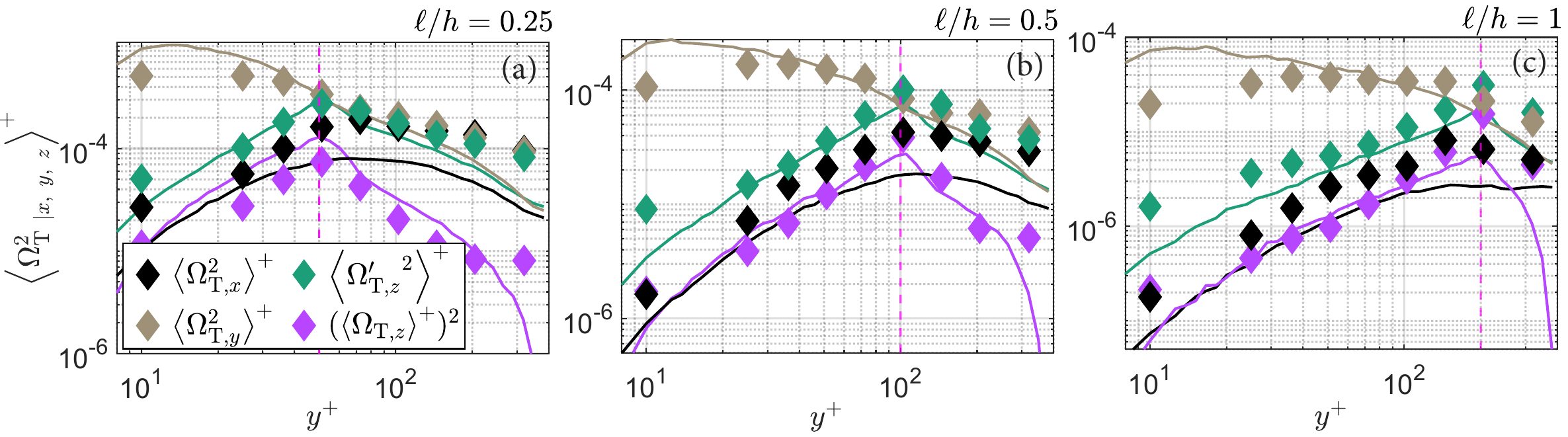}}
  \caption{Decomposition of the mean-square tumbling rate~$\langle \boldsymbol{\Omega}_{\rm T}^2 \rangle$ into the four different components of equation~\eqref{eq:tumbling_decomposition}, for the different fibre lengths: $\ell/h=0.25$ (a), $\ell/h=0.5$ (b) and $\ell/h=1$ (c). Experimental data are represented by diamonds, while numerical simulations are shown through solid lines of corresponding colours. }
\label{fig:Omega}
\end{figure}

We see from figure~\ref{fig:tumbrate} that the tumbling rate decreases systematically with fibre length. This trend is qualitatively consistent with homogeneous isotropic turbulence, where, for fibres whose length lies in the inertial range, $\langle|\dot{\boldsymbol{p}}|^2\rangle\tau_K^2 \propto(\ell/\eta_K)^{-4/3}$ \citep{parsa2014,oehmke}. This scaling need not apply directly to the present wall-bounded flow, which is both inhomogeneous and anisotropic. To identify the contributions responsible for the observed profiles, we decompose the tumbling angular velocity into its laboratory-frame components. Writing $\Omega_{{\rm T},i} =\langle\Omega_{{\rm T},i}\rangle+\Omega_{{\rm T},i}'$ gives $\langle\Omega_{{\rm T},i}^2\rangle =\langle\Omega_{{\rm T},i}\rangle^2+\langle\Omega_{{\rm T},i}'^2\rangle$. Only the mean spanwise component, $\langle\Omega_{{\rm T},z}\rangle$, is appreciably non-zero, as expected from the mean streamwise shear. The mean streamwise and wall-normal components are negligible. The total mean-square tumbling rate can therefore be written as
\begin{equation}
   \langle \boldsymbol{\Omega}_{\rm T}^2 \rangle = \langle \Omega_{\rm T,x}'^2 \rangle + \langle \Omega_{\rm T,y}'^2 \rangle +\langle \Omega_{\rm T,z}'^2 \rangle + \langle \Omega_{\rm T,z} \rangle^2, \label{eq:tumbling_decomposition}
\end{equation}
with $\langle \Omega_{\rm T,x}'^2 \rangle = \langle \Omega_{\rm T,x}^2 \rangle$ and $\langle \Omega_{\rm T,y}'^2 \rangle = \langle \Omega_{\rm T,y}^2 \rangle$. The four terms of this decomposition are shown in figure~\ref{fig:Omega} for the three fibre lengths.

For all fibre lengths, $\langle\Omega_{{\rm T},y}'^2\rangle$ is the dominant contribution close to the wall. It forms a plateau in the experiments and decreases slowly with wall distance in the simulations. This component is associated with kayaking events, such as that shown in figure~\ref{fig:3dRecon}(b), during which the fibre rotates about the wall-normal axis in a plane parallel to the wall. Because this is the dominant near-wall contribution, it accounts for most of the difference between the experimental and numerical total tumbling rates in this region. It begins to decrease more rapidly slightly below $y^+=\ell^+/2$. The fluctuating spanwise contribution $\langle\Omega_{{\rm T},z}'^2\rangle$ is small very close to the wall but increases with wall distance and reaches a maximum near $y^+=\ell^+/2$, where it becomes comparable to $\langle\Omega_{{\rm T},y}'^2\rangle$. It therefore accounts for the peak in the total tumbling rate. We associate this maximum with pole-vaulting events, during which a nearly wall-normal fibre undergoes strong rotation after one end touches the wall; an example is shown in figure~\ref{fig:3dRecon}(a). Beyond the maximum, $\langle\Omega_{{\rm T},z}'^2\rangle$ decreases with wall distance and remains comparable in magnitude to the wall-normal contribution. The mean-shear contribution $\langle\Omega_{{\rm T},z}\rangle^2$ has a similar wall-normal dependence but is approximately four to five times smaller. It decreases more rapidly beyond $y^+=\ell^+/2$, indicating that the direct contribution of the mean shear is significant only near the wall. The streamwise contribution~$\langle\Omega_{{\rm T},x}'^2\rangle$ is the smallest at all wall distances and for all fibre lengths, although it also exhibits a weak maximum near $y^+=\ell^+/2$. Similar component-wise trends were reported by \citet{baker_coletti_2022} for a single fibre length. In particular, their data also show a maximum of $\langle\Omega_{{\rm T},z}'^2\rangle$ near $y^+\simeq\ell^+/2\simeq30$, although this connection with fibre length was not emphasised.

\begin{figure}
  \centerline{\includegraphics[width=.85\textwidth]{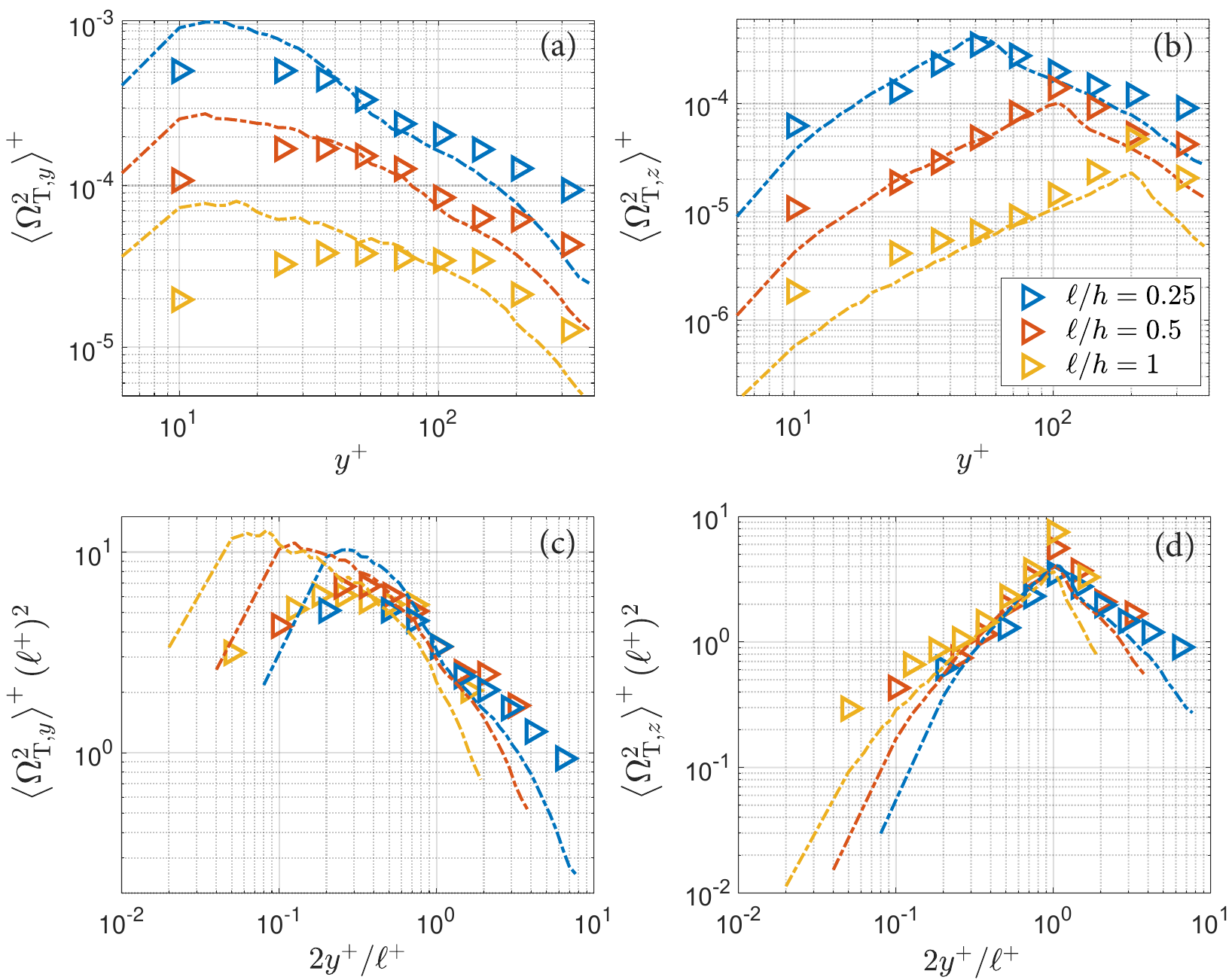}}
  \caption{Wall-normal (a,c) and spanwise (b,d) components of the mean-square tumbling rate for the three fibre lengths. The top row shows the unscaled profiles as a function of $y^+$. The bottom row shows the compensated quantities $\langle\Omega_{{\rm T},i}^{+\,2}\rangle(\ell^+)^2$ as functions of $2y^+/\ell^+$, so that the geometrical location $y^+=\ell^+/2$ corresponds to $2y^+/\ell^+=1$.}
\label{fig:Omeganorm}
\end{figure}

We now examine the length dependence of the two dominant components. The spanwise component $\langle\Omega_{{\rm T},z}^2\rangle$, shown in figure~\ref{fig:Omeganorm}(b), exhibits a clear maximum near $y^+=\ell^+/2$. During a pole-vaulting event, the fibre is nearly wall-normal and touches the wall at one end. Taking $u_\tau$ as the characteristic velocity and $\ell$ as the characteristic lever arm gives $\langle\Omega_{{\rm T},z}^2\rangle\sim\left(u_\tau/\ell\right)^2$. Equivalently, defining $\Omega_{\rm T}^+=\tau_\nu\Omega_{\rm T}$ gives $\langle\Omega_{{\rm T},z}^{+\,2}\rangle\sim(\ell^+)^{-2}$. The compensated profiles $\langle\Omega_{{\rm T},z}^{+\,2}\rangle(\ell^+)^2$, plotted in figure~\ref{fig:Omeganorm}(d), approximately collapse when expressed as a function of the rescaled wall distance. This supports the proposed $(\ell^+)^{-2}$ scaling near the pole-vaulting maximum. The experimental maximum is sharper than its numerical counterpart, and this difference increases with fibre length. Possible effects of gravity and of the fibre--wall collision model are discussed in \S\ref{sec:discussion}.

The same compensation also produces an approximate collapse of the wall-normal component $\langle\Omega_{{\rm T},y}^{+\,2}\rangle$ in the vicinity of $y^+=\ell^+/2$ [figure~\ref{fig:Omeganorm}(c)]. The quality of the collapse deteriorates closer to the wall, indicating that the same dimensional compensation does not necessarily reflect the same dynamical mechanism in both regions. Near the wall, the dominant wall-normal component is associated with kayaking, namely rotation about the wall-normal axis in a plane parallel to the wall, rather than with pole vaulting.

\section{Discussion on the comparison between experiments and simulations}
\label{sec:discussion}

The results presented in \S\ref{sec:results} show that the simulations based on slender-body theory reproduce most of the experimental trends, including their dependence on fibre length and wall distance. The remaining discrepancies are nevertheless systematic and provide information on the physical effects that are not represented in the numerical model. We organise their discussion around three main modelling differences: \textit{(i)}~the absence of gravity, \textit{(ii)}~the simplified treatment of fibre--wall interactions, and \textit{(iii)}~the finite diameter of the experimental fibres. These effects are not completely independent: gravity controls how frequently fibres encounter the bottom wall, the collision model determines their dynamics during such encounters, and the finite diameter modifies both near-wall interactions and the hydrodynamic forces and torques exerted by the fluid. The following interpretation therefore relies on the consistency between several independent observables rather than on a one-to-one attribution of each discrepancy.

\subsection{Effects of gravity and settling-induced sampling}
\label{sec:discussion_gravity}

The most direct consequence of gravity is the strong asymmetry of the experimental concentration profile shown in figure~\ref{fig:density}(a). The negatively buoyant fibres preferentially accumulate near the bottom wall, whereas the gravity-free simulations yield statistically equivalent populations near the two walls [figure~\ref{fig:density}(b)]. More importantly, sedimentation affects not only the spatial distribution of the fibres but also the turbulent events that they must sample in order to reach and remain at a given wall distance. The experimental fibres observed away from the bottom wall do not constitute an unbiased sample of the suspension: many of them have recently been transported out of the near-wall region by sufficiently strong upward motions.

Finite particle inertia can itself lead to preferential sampling of turbulent structures, for instance through an under-sampling of vortical regions and an over-sampling of strain-dominated regions~\citep{squires,Jeremie2023}. The leading-order inertial response responsible for this mechanism is represented in both the experiments and the simulations and therefore cannot, by itself, explain their differences. Its role here is instead to provide the finite response time through which the gravity-induced selection of trajectories leaves a persistent signature on the fibre statistics. The discrepancies discussed below are therefore attributed primarily to the combined effects of settling-induced sampling and inertial memory.

This settling-induced selection provides a possible interpretation of the experimental streamwise velocity lag. A fibre lifted from the near-wall region retains, over its finite response time, part of the lower streamwise velocity acquired closer to the wall. It can therefore reach the channel bulk with a streamwise velocity smaller than the local mean fluid velocity. This history effect may explain why the experimental fibres lag the fluid in figures~\ref{fig:velocity_mean}(a) and~\ref{fig:vpdfexp}(e), whereas no comparable deficit is obtained in the gravity-free simulations. Conditioning the statistics on the instantaneous wall distance does not remove such a bias because it does not condition on the history of the fibre trajectory.

The wall-normal velocity statistics provide evidence that the experimental fibres do not sample upward and downward motions symmetrically. Because upward motions must oppose gravitational settling for fibres to remain away from the bottom wall, the observed population is biased towards sufficiently intense positive wall-normal velocities. This mechanism is consistent with the weak overall asymmetry of the experimental wall-normal velocity distributions towards positive values [figures~\ref{fig:vpdfexp}(b),~\ref{fig:vpdfexp}(d) and~\ref{fig:vpdfexp}(f)]. The more pronounced positive tails observed in specific wall-normal intervals are instead associated with pole-vaulting events and are discussed separately below.

The fibre Reynolds shear stress nevertheless shows that this upward selection cannot be identified with an enhanced sampling of canonical low-speed ejections. Its magnitude is smaller in the experiments than in the simulations [figure~\ref{fig:velocity_mean}(b)]. The experimental streamwise lag and positive wall-normal skewness therefore cannot be explained simply by a stronger instantaneous sampling of events with $u_p'<0$ and $v_p'>0$, which would instead make $\langle u_p'v_p'\rangle$ more negative. The Reynolds stress depends both on the occurrence of upward motions and on the streamwise-velocity regions sampled during these motions. Gravity and fibre--wall interactions may promote upward transport from a broader range of near-wall environments, including high-speed regions and interfaces between streaks, where $u_p'$ is only weakly negative or can be positive. This possibility is consistent with the length-dependent sampling of high- and low-speed near-wall regions inferred from figure~\ref{fig:vpdfexp}(a). Inertial memory may further weaken the instantaneous correlation: a fibre transported away from the wall can retain a streamwise velocity deficit after its wall-normal velocity has decorrelated from the event that initially lifted it. The mean streamwise lag, positive wall-normal skewness, and reduced magnitude of $\langle u_p'v_p'\rangle$ are therefore not contradictory, but probe different aspects of the fibre trajectories.

Gravity may similarly account for part of the discrepancy in the rotational statistics. In the experiments, the tumbling rate decreases more slowly towards the channel centre than in the simulations, as shown in figure~\ref{fig:tumbrate}. Fibres observed far from the bottom wall have typically been transported from the near-wall region by sufficiently energetic upward motions. Since the mean tumbling rate is larger close to the wall, they may retain, over their finite rotational response time, part of the enhanced angular velocity acquired there. The turbulent events responsible for their transport may additionally involve strong velocity gradients and further increase their rotational activity. By contrast, fibres at the same wall distance in the gravity-free simulations are not selected according to their ability to overcome settling. The experimental bulk population can consequently display a larger tumbling rate even when the local carrier-flow statistics are comparable.

Sedimentation may also increase the occurrence of pole-vaulting by continuously driving fibres towards the bottom wall. This could contribute to the sharper experimental maximum of the spanwise tumbling rate near $y^+\simeq\ell^+/2$, visible in figures~\ref{fig:Omeganorm}(b) and~\ref{fig:Omeganorm}(d). Gravity alone, however, is unlikely to determine the detailed amplitude and width of this peak, which also depend on the dynamics of the fibre during contact with the wall and are therefore discussed in \S\ref{sec:discussion_wall}.

The role of gravity could be isolated experimentally by varying the density contrast while keeping the fibre geometry and the carrier flow as unchanged as possible. Such measurements would determine whether the bulk velocity lag, wall-normal skewness, Reynolds shear stress, and enhanced tumbling vary consistently with the settling velocity.

\subsection{Fibre--wall interactions and collision modelling}
\label{sec:discussion_wall}

The limitations of the wall model become particularly apparent when gravity is included in the simulations. The buoyancy-corrected gravitational acceleration was chosen to reproduce the settling velocity measured for the experimental fibres in still water. Nevertheless, for all the gravitational accelerations considered, the simulated fibres eventually sediment onto the bottom wall and remain there. The simulations therefore fail to reproduce the statistically stationary experimental state in which negatively buoyant fibres repeatedly encounter the wall but are also returned to the flow. This failure indicates that an important mechanism is missing from the model of near-wall hydrodynamics or contact dynamics.

In the present simulations, wall interactions are represented by a simple elastic collision law. In particular, the model does not include lubrication forces or tangential contact forces. Lubrication becomes important when the gap between a finite-diameter fibre and the wall is small compared with the fibre diameter. It strongly resists normal approach and can substantially increase the duration of a near-wall interaction. Tangential friction can additionally constrain sliding of the point of contact and redistribute translational and rotational momentum. Neither effect is represented by an instantaneous frictionless collision of an infinitely thin centreline. Lubrication alone should not be regarded as a repulsive mechanism capable of resuspending a deposited fibre, but its combination with tangential contact forces and turbulent forcing may strongly modify the sequence of approach, impact, sliding, pivoting, and detachment.

Several orientation and rotation statistics support the importance of these effects. For the longer fibres, the simulations show a bimodal distribution of the wall-normal orientation component $p_y$ in the logarithmic layer, whereas the two peaks are not resolved experimentally [figure~\ref{fig:OrienPDFall}(e)]. Within the idealised collision model, the approach to and departure from the wall can produce relatively distinct nose-down and nose-up configurations. A finite-duration contact involving lubrication, sliding resistance, or pivoting about an approximately fixed contact point would instead allow the fibre to explore a broader range of intermediate orientations and could smooth this bimodality. This interpretation is necessarily qualitative, but the discrepancy is localised in precisely the configurations for which contact dynamics should be most influential.

The near-wall differences in mean streamwise velocity [figures~\ref{fig:velocity_mean}(a) and~\ref{fig:vpdfexp}(a)] and kayaking rate [figure~\ref{fig:Omega}] may have the same origin. The wall interaction controls whether a fibre slides freely along the wall, pivots about one end, or rapidly rebounds into the flow. These different motions determine both the residence time of the fibre close to the wall and its ability to remain within, or move across, low- and high-speed streaks. An imperfect description of contact can therefore alter the streamwise velocity sampled by the fibre as well as the relative amplitudes of the wall-normal and spanwise tumbling components associated with kayaking and pole-vaulting. This contact-related contribution is distinct from the modification of the hydrodynamic coupling to near-wall structures caused by the finite fibre diameter, discussed in \S\ref{sec:discussion_diameter}.

The pole-vaulting peaks provide a particularly direct diagnostic. Their position near $y^+=\ell^+/2$ has a simple geometrical interpretation: it corresponds to fibres whose centre is approximately half a fibre length from the wall while one end interacts with it. The sharper and more pronounced experimental peaks in figures~\ref{fig:Omega},~\ref{fig:Omeganorm}(b) and~\ref{fig:Omeganorm}(d) suggest that these constrained pivoting configurations persist longer, or occur more coherently, than in the simulations. The elastic collision model appears to capture the geometry of pole-vaulting but not its complete dynamics. Gravity can enhance the rate at which fibres enter such configurations, as discussed above, whereas lubrication and tangential contact forces are expected to control their duration and rotational signature.

A more realistic wall model should therefore distinguish between normal lubrication, direct contact, tangential sliding, and pivoting. Testing such a model in simulations with gravity would provide a stringent criterion: it should permit a statistically stationary balance between deposition and resuspension while simultaneously improving the concentration profile, the statistics of $p_y$, the near-wall translational velocity, and the pole-vaulting and kayaking rates.

\subsection{Finite-diameter effects and limitations of slender-body theory}
\label{sec:discussion_diameter}

The experimental fibres have a finite diameter of approximately $d^+\simeq10$. The Reynolds number based on friction velocity is thus $\Rey_d=u_\tau d/\nu\simeq 10$. Since the characteristic fibre--fluid slip velocity is a non-negligible fraction of $u_\tau$, the corresponding slip-based Reynolds number, $\Rey_d^{\rm slip}=U_{\rm slip}d/\nu$, is expected to be of order unity or larger. The considered fibres therefore do not strictly satisfy either the asymptotically small slenderness approximation or the vanishing-fluid-inertia assumption underlying the present slender-body model. Finite-diameter effects can influence the translational dynamics through nonlinear, orientation-dependent drag and the rotational dynamics through inertial torque corrections. They can also modify how fibres interact with near-wall coherent structures even in the absence of direct contact.

The discrepancy in the near-wall streamwise velocity, shown in figures~\ref{fig:velocity_mean}(a) and~\ref{fig:vpdfexp}(a), is consistent with such an interpretation. Previous numerical studies based on point-particle or slender-body descriptions have reported preferential sampling of low-speed streaks and no corresponding excess of particle velocity near the wall~\citep{mortensen2008678,marchioli2010}. By contrast, the interface-resolved simulations of \citet{duo} found that finite-size fibres can preferentially accumulate in high-speed streaks. The difference between the present simulations and experiments may therefore reflect more than the collision law alone: resolving the finite fibre cross-section can change the hydrodynamic coupling to streaks and hence the streamwise velocity sampled during near-wall motion.

Finite-diameter effects may also be involved in the anomalous spanwise alignment of the two longer experimental fibre types within approximately $30\lesssim y^+\lesssim90$, shown in figures~\ref{fig:Orientation}(b) and~\ref{fig:Orientation}(c). The spanwise spacing of near-wall coherent structures is of the order of $100$ wall units~\citep{adrian_2007}. The two longest fibres are comparable to, or longer than, this spacing and can therefore simultaneously sample velocity differences across neighbouring structures~\citep{shaik1}. Fibre length alone cannot explain the discrepancy, since this non-local sampling is already accounted for in the slender-body simulations. However, at finite~$\Rey_d$, the force and torque generated by such spatially varying slip velocities need not coincide with their linear Stokesian approximations. Inertial corrections could consequently increase the persistence of orientations transverse to the streamwise direction.

\begin{figure}
\centerline{\includegraphics[width=.45\textwidth]{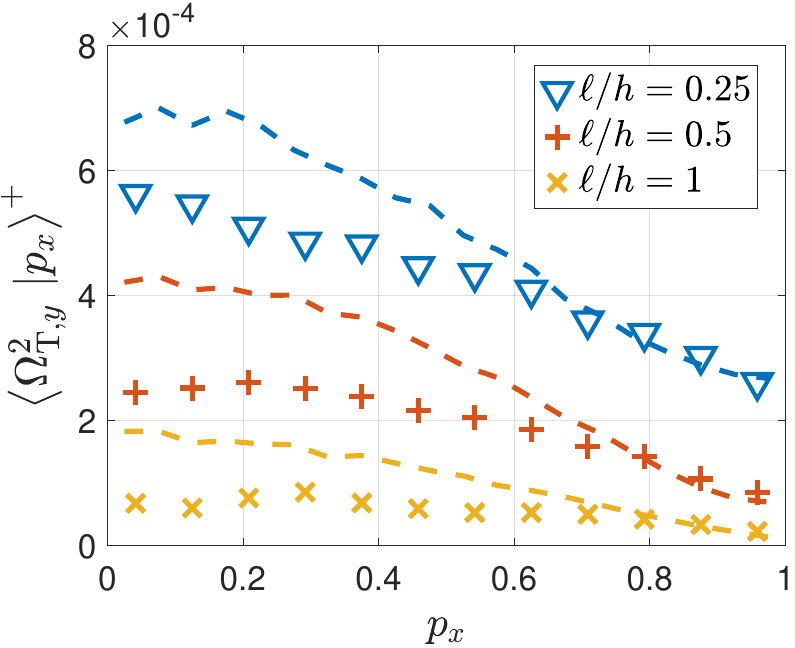}}
\caption{Wall-normal component of the mean-square tumbling angular velocity, conditioned on the streamwise orientation component $p_x$, for the different fibre lengths and for fibre centres located within $30\leq y^+\leq60$.}
\label{fig:OrienCorrel}
\end{figure}
Figure~\ref{fig:OrienCorrel} provides a more discriminating test of this possibility. It shows the wall-normal component of the mean-square tumbling angular velocity, $\langle\Omega_{\mathrm{T},y}^2\rangle$, conditioned on the streamwise orientation component $p_x$ for fibres with centres in $30\leq y^+\leq60$. For all fibre lengths, the tumbling rate decreases as $p_x$ increases: fibres aligned with the streamwise direction rotate more slowly and consequently remain aligned for longer. For $p_x\simeq1$, experiments and simulations agree well, indicating that the model captures the rotational dynamics of strongly streamwise-aligned fibres. The agreement progressively deteriorates as $p_x$ decreases, and the discrepancy increases markedly with fibre length. It remains of the order of $15\%$ for the shortest fibres but exceeds $100\%$ for the longest ones.

The simulations therefore indicate that long fibres in non-streamwise orientations rotate much faster than they do experimentally. The excess spanwise alignment observed in the experiments is consistent with this reduced tumbling: once a long experimental fibre reaches a non-streamwise, and in particular a spanwise-oriented, configuration, it remains there for longer than predicted by the slender-body model. This conditioned statistic therefore points to a missing orientation-dependent rotational mechanism. In particular, it shows that the orientation discrepancy does not arise solely from differences in the concentration profile or from a uniform error in the tumbling rate; it is associated specifically with the dynamics of long fibres away from streamwise alignment.

At finite $\Rey_d^{\rm slip}$, both the drag and the hydrodynamic torque can acquire nonlinear corrections that depend on the instantaneous orientation and slip velocity. Such corrections offer a possible common origin for the near-wall streamwise-velocity discrepancy and for the excessive tumbling of non-streamwise fibres in the simulations. An inertial extension of slender-body theory applicable up to $\Rey_d^{\rm slip}\simeq10$ has recently been developed by \citet{Joshietal2026}. Implementing this formulation would allow the influence of fluid inertia to be assessed without immediately resorting to fully interface-resolved simulations.

Experimentally, varying the fibre diameter would provide the most direct test of this interpretation, but it would simultaneously change the translational and rotational response times, the settling velocity, and the wall-contact dynamics. Numerically, inertial slender-body calculations followed by selected interface-resolved simulations may therefore provide a more controlled route for separating finite-diameter hydrodynamic effects from gravity and wall-collision effects.

\section{Conclusions}\label{sec:ccl}

In this work, we have investigated the dynamics of long rigid fibres in a turbulent channel flow at $\Rey_\tau \approx 400$ through a systematic comparison between experiments and simulations. Three fibre lengths were considered, all much larger than the viscous length scale of the flow, $\ell^+\approx100$, $200$, and $400$, and representing substantial fractions of the channel half-height, $\ell/h=0.25$, $0.5$, and $1$. Our results show that fibre length is a central control parameter of the dynamics. In particular, finite-size confinement and interactions with the walls strongly affect fibre translation, orientation, and rotation.

In the experiments, the fibres lag behind the fluid away from the wall, with no clear dependence of the mean velocity deficit on fibre length. The behaviour is more strongly length-dependent  near the wall: the shortest fibres move, on average, faster than the local mean fluid velocity, whereas the longer fibres continue to lag behind it. This difference is associated with the preferential sampling of distinct regions of the near-wall flow, and in particular of high- or low-speed streaks. By contrast, the numerical fibres display only weak departures from the mean fluid velocity. This constitutes the most significant discrepancy between experiments and simulations and is likely related primarily to the omission of gravity in the simulations used to obtain statistically stationary suspension statistics.

The orientation statistics are in comparatively good agreement between experiments and simulations. The mean-square components of the fibre orientation collapse reasonably well when the wall distance is normalised by the fibre length, showing that $y^+/\ell^+$ is the relevant geometrical variable for describing the transition from wall-induced alignment to bulk behaviour. The remaining discrepancies, most visible over the range $30\lesssim y^+\lesssim 90$, are consistent with finite-diameter and finite-slip-Reynolds-number effects that are not represented in the numerical slender-body model.

A similarly good overall agreement is obtained for the tumbling statistics. The tumbling rate reaches a maximal intensity at a wall distance of order $y^+=\ell^+/2$, corresponding to the region in which geometrical confinement and wall collisions become important. The tumbling rate magnitude decreases markedly with fibre length, as previously observed in homogeneous and isotropic turbulence, although with a different length dependence. Near the maximum, the tumbling intensity is rescaled by $(\ell^+)^2$. This scaling is consistent with a dimensional estimate based on the velocity variation sampled by a fibre within the boundary layer. These results demonstrate that, for fibres whose length is comparable to the characteristic wall-normal scales of the flow, their rotational dynamics cannot be understood solely from local velocity-gradient statistics.

The remaining differences between experiments and simulations point towards several extensions of the present work. A more complete numerical description should incorporate gravitational settling together with finite-diameter, finite-slip-Reynolds-number and wall-contact effects, while maintaining a statistically stationary suspension. Fibre flexibility is another natural direction, since sufficiently slender fibres can deform under the strong and spatially heterogeneous hydrodynamic stresses encountered near the wall. Finally, the substantial settling velocity of the experimental fibres suggests connections with sediment transport, including saltation, deposition and resuspension. The coupling between these processes and the orientation-dependent dynamics of anisotropic particles remains largely unexplored and may lead to regimes qualitatively different from those observed for spherical particles.

\backsection[Acknowledgements]{Gautier Verhille is warmly acknowledged for having shared his 3D fibre reconstruction codes and his knowledge on how to use and adapt them to our setup. Kévin Gutierrez is thanked for having performed preliminary experiments on the channel flow setup.}

\backsection[Funding]{This work received support from the UCA-JEDI Future Investments funded by the French government (Grant No. ANR-15-IDEX-01) and from the Agence Nationale de la Recherche (Grants No. ANR-21-CE30-0040-01 and No. ANR-23-CE30-0005-01). Computational resources were provided by the OPAL infrastructure from Université Côte d'Azur.}

\backsection[Declaration of interests]{The authors report no conflict of interest.}


\backsection[Author ORCIDs]{D. Sun, https://orcid.org/0009-0008-0115-3291; C. Claudet. https://orcid.org/0000-0002-7733-8283; J. Bec, https://orcid.org/0000-0002-3618-5743; C. Brouzet, https://orcid.org/0000-0003-3131-3942} 


\bibliographystyle{jfm}

\end{document}